\documentclass[%
amsmath,amssymb,aps,
superscriptaddress,
preprintnumbers,
nofootinbib, cases,
 amsmath,amssymb,
 aps,
longbibliography,
]{revtex4-2}

\pdfoutput=1
\usepackage{aas_macros}
\usepackage[sort&compress]{natbib}
\usepackage{url}
\usepackage{hyperref}
\usepackage{amsfonts}% Include figure files
\usepackage{amsmath}
\usepackage{physics}
\usepackage{epsfig}\usepackage{dcolumn}% Align table columns on decimal point
\usepackage{amssymb}
\usepackage{cleveref}
\usepackage{booktabs}
\usepackage{bm}% bold math
\usepackage{slashed}
\graphicspath{{./figures/}}
\usepackage{subcaption}
\usepackage{graphicx}
\usepackage{physics}
\usepackage{cancel}
\usepackage{multirow}

\newcommand{\ee}[1]{\times 10^{#1}}

\newcommand{\sv}{\langle \sigma v \rangle}

\usepackage{dcolumn}% Align table columns on decimal point
\usepackage{bm}% bold math
\usepackage{hyperref}
\usepackage{cleveref}
\usepackage{booktabs}
\usepackage{physics}
\usepackage[english]{babel}
\usepackage[dvipsnames]{xcolor}

\usepackage{epsfig}
\usepackage{amssymb}
\usepackage{amsmath}
\usepackage{hyperref}
\usepackage{xcolor}
\usepackage{subcaption}
\usepackage{graphicx}
\usepackage{wrapfig}
\usepackage{dblfloatfix}
\usepackage[normalem]{ulem}
\begin{document}

%%%%%%%%%%%%%%%%%%%%%%%%%%%%%%%%%%%%%%%%%%%%%%%%%%%%

\title{Hunting for Dark Matter and New Physics with GECCO}

\author{Adam Coogan}%
\email{adam.coogan@umontreal.ca}
\affiliation{Gravitation Astroparticle Physics Amsterdam (GRAPPA),\\ Institute for Theoretical Physics Amsterdam and Delta Institute for Theoretical Physics,\\ University of Amsterdam, Science Park 904, 1098 XH Amsterdam, The Netherlands}
\affiliation{Département de Physique, Université de Montréal, 1375 Avenue Thérèse-Lavoie-Roux, Montréal, QC H2V 0B3, Canada}
\affiliation{Mila -- Quebec AI Institute, 6666 St-Urbain, \#200, Montreal, QC, H2S 3H1}

\author{A.A. Moiseev}%
\email{amoiseev@umd.edu}
\affiliation{University of Maryland, College Park, MD 20742, and CRESST/NASA/Goddard Space Flight Center, Greenbelt, MD 20771, USA}

\author{Logan Morrison}%
\email{loanmorr@ucsc.edu}
\affiliation{Department of Physics, University of California, Santa Cruz, CA 95064, USA}
\affiliation{Santa Cruz Institute for Particle Physics, Santa Cruz, CA 95064, USA}

\author{Stefano Profumo}%
\email{profumo@ucsc.edu}
\affiliation{Department of Physics, University of California, Santa Cruz, CA 95064, USA}
\affiliation{Santa Cruz Institute for Particle Physics, Santa Cruz, CA 95064, USA}

%%%%%%%%%%%%%%%%%%%%%%%%%%%%%%%%%%%%%%%%%%
\author{Matthew~G.~Baring} % [0000-0003-4433-1365]
\affiliation{Department of Physics and Astronomy - MS 108,
Rice University, 6100 Main Street, Houston, Texas 77251-1892, USA}

\author{Aleksey Bolotnikov}
%\email{vigliano.alessandroarmando@spes.uniud.it}
\affiliation{Brookhaven National Laboratory, Upton, NY 11973, USA}

\author{Gabriella A. Carini}
\affiliation{Brookhaven National Laboratory, Upton, NY 11973, USA}

\author{Sven C. Herrmann}
\affiliation{Brookhaven National Laboratory, Upton, NY 11973, USA}

\author{Francesco Longo}
%\email{francesco.longo@ts.infn.it}
\affiliation{Department of Physics, University of Trieste, Trieste, Italy}
\affiliation{Istituto Nazionale di Fisica Nucleare (INFN), sezione di Trieste, Trieste, Italy}

\author{Floyd W. Stecker}
%\email{Floyd.W.Stecker@nasa.gov}
\affiliation{Astrophysics Science Division, NASA Goddard Space Flight Center, Greenbelt, MD 20771, USA}
\affiliation{Department of Physics and Astronomy,  University of California, Los Angeles, CA 90095, USA}

\author{Alessandro Armando Vigliano}
%\email{vigliano.alessandroarmando@spes.uniud.it}
\affiliation{Department of Mathematical, Computer and Physical Sciences, University of Udine, Udine, Italy}
\affiliation{Istituto Nazionale di Fisica Nucleare (INFN), sezione di Trieste, Trieste, Italy}

\author{Richard S. Woolf}
%\email{richard.woolf@nrl.navy.mil}
\affiliation{Space Science Division, U.S. Naval Research Laboratory, Washington, DC, USA}

\date{\today}

\begin{abstract}
%%%%%%%%%%%%%%%%%%%%%%%%%%%%%%%%%%%%%%%%%%%%%%%%%%%%
We outline the science opportunities in the areas of searches for dark matter and new physics offered by a proposed future MeV gamma-ray telescope, the Galactic Explorer with a Coded Aperture Mask Compton Telescope (GECCO). We point out that such an instrument would play a critical role in opening up a discovery window for particle dark matter with mass in the MeV or sub-MeV range, in disentangling the origin of the mysterious 511 keV line emission in the Galactic Center region, and in potentially discovering Hawking evaporation from light primordial black holes.
\end{abstract}

%%%%%%%%%%%%%%%%%%%%%%%%%%%%%%%%%%%%%%%%%%%%%%%%%%%%

\maketitle
\clearpage

%%%%%%%%%%%%%%%%%%%%%%%%%%%%%%%%%%%%%%%%%%%%%%%%%%%%%%%%%%%
\section{Introduction}
%%%%%%%%%%%%%%%%%%%%%%%%%%%%%%%%%%%%%%%%%%%%%%%%%%%%

%The ``MeV gap''
It is in not an overstatement that the MeV gamma-ray energy range remains one of the least explored frontiers in observational astronomy, with important implications for the understanding of high-energy astrophysical phenomena. With the most recent data dating back several decades, the photon band in between hard x-rays and the gamma rays detectable with the Fermi Large Area Telescope offers some of the richest opportunities for discovery across the electromagnetic spectrum. It is therefore not a surprise that much activity has resumed in recent years around a next-generation MeV telescope. Without attempting to be exhaustive, a partial list of such missions under consideration, in no special order, includes AdEPT~\cite{adept}, AMEGO~\cite{amego}, e\-ASTROGAM~\cite{e_astrogam, as_astrogam},
%\footnote{This has since been scaled back to All-Sky-ASTROGAM~\cite{as_astrogam}.},
MAST~\cite{mast}, COSI~\cite{Tomsick}, PANGU~\cite{pangu,pangu_aeff} and GRAMS~\cite{grams,grams_loi}.

The scientific significance of a new space-borne observatory in the MeV range includes a very broad range of topics such as identifying the hadronic versus leptonic nature and the acceleration processes underpinning jet outflows, studying the role of magnetic fields in powering the jets associated with gamma-ray bursts, pinning down the sources of gravitational wave events, and  understanding the electromagnetic counterparts of astrophysical neutrinos. Lower energy phenomena will also be clarified by new capabilities in the MeV: for instance, cosmic-ray diffusion in interstellar clouds, and the role cosmic rays play in gas dynamics and wind outflows, as well as nucleosynthesis and chemical enrichment via the study of nuclear emission lines.

Here, we focus on a proposed mid-size ``Explorer'' (MIDEX) class mission, the Galactic Explorer with a Coded Aperture Mask Compton Telescope (GECCO) \cite{Orlando:2021get} and consider its capabilities in the search for new physics beyond the Standard Model. We describe GECCO in some detail in the following \cref{sec:gecco}. We then explore GECCO's potential in searching for dark matter annihilation and decay for dark matter particle masses in the MeV range in \cref{sec:dm}; in discovering the products of Hawking evaporation of primordial black holes in \cref{sec:pbh} (see also Ref.~\cite{Coogan:2020tuf}); and in identifying the origin of the 511 keV emission line from the Galactic Center (\cref{sec:511}).

%%%%%%%%%%%%%%%%%%%%%%%%%%%%%%%%%%%%%%%%%%%%%%%%%%%%%%%%%%%
\section{The Galactic Explorer with a Coded Aperture Mask Compton Telescope}
\label{sec:gecco}

The Galactic Explorer with a Coded Aperture Mask Compton Telescope (GECCO) is a novel concept for a next-generation $\gamma$-ray telescope that will cover the hard x-ray to soft $\gamma$-ray region, and is currently being considered for a future NASA Mid-Size Explorer (MIDEX) class mission \cite{gecco1, Orlando:2021get}. 
GECCO will conduct high-sensitivity measurements of the cosmic $\gamma$-radiation in the energy range from 50 keV to $\sim10$ MeV and create intensity maps with high spectral and spatial resolution, with a focus on the separation of diffuse and point-source components. Its science objectives are focused on understanding the nature, composition and fine structure of the inner Galaxy, on the discernment of the origin of the positron annihilation 511 keV line, identification and precise localization of gravitational wave and neutrino events, and on the resolution of the Galactic chemical evolution and sites of explosive elements synthesis by precise measurements of nuclear lines topography. As we show in this study, GECCO’s observational capabilities will be of paramount importance for e.g. disentangling astrophysical and dark matter explanations of emission from the Galactic Center and potentially providing a key to discovering as-of-yet unexplored dark matter candidates \cite{gecco_loi, Orlando:2021get}.

\begin{figure*}
    \centering
    \includegraphics[width=1.0\textwidth]{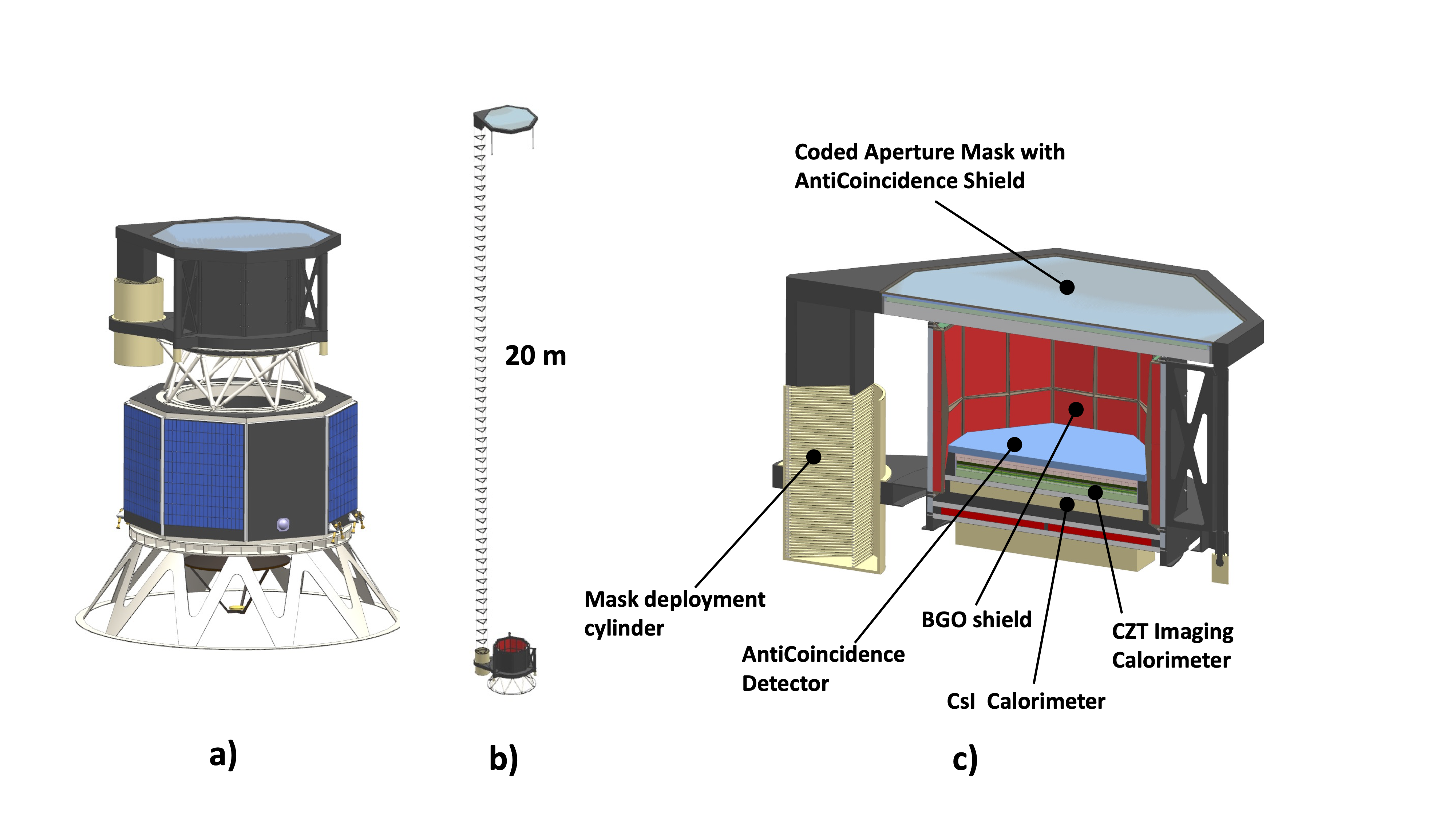}
    \caption{GECCO design concept: a) with mask in stowed position and notional spacecraft bus; b) with mask in deployed position; c) cutaway. \cite{gecco1}}
    \label{fig:gecco-concept}
\end{figure*}

\subsection{Instrument concept}

GECCO is a modern $\gamma$-ray telescope  designed according to two combined principles:  Compton imaging and coded-aperture mask imaging. This combination mutually enhances the performance of each telescope and enables previously inaccessible measurements. Compton telescopes provide good, low-noise performance and allow for a wide field-of-view (FoV), but Doppler broadening fundamentally limits the achievable angular resolution to $\sim$1 degree. Conversely, coded aperture telescopes can achieve very high angular resolution at arcmin level in point source detection and localization, but are unable to detect diffuse radiation, have limited FoV and practically no inherent background rejection. Combining a coded aperture mask with an imaging detector that is also a Compton telescope will widen the potential scope of the instrument objectives.  Given the scope of this paper, we will address only the high angular resolution measurements with coded-aperture mask and the measurements sensitivity (see \cite{gecco1,Orlando:2021get} for GECCO details). The combination of a coded aperture mask with a Compton telescope has been previously demonstrated in simulations \cite{Galloway,Aprile}, and tested with INTEGRAL/IBIS data \cite{Forot}, but the mature concept has never been implemented as the central motivation for a telescope design.

GECCO has an octagon shape with a medium diagonal of $\sim$90 cm. The instrument is based on a novel Cadmium-Zinc-Telluride (CZT) imaging calorimeter and a deployable coded aperture mask. It also utilizes a heavy-scintillator (BGO) shield, a CsI calorimeter, and a plastic scintillator anticoincidence detector (fig.\ref{fig:gecco-concept}). The CZT Imaging Calorimeter detects incident photons in an energy range from  $\sim$100 keV to  $\sim$10 MeV with $>50$\% efficiency, measuring points of photon interaction with 3D accuracy better than 1mm and deposited energy with 1-2\% FWHM (full width half maximum) resolution. The base element of the calorimeter is a virtual Frisch grid drift CZT bar with the baseline dimensions 8mm x 8mm x 32mm, where the coordinates of the photon interaction are measured, along with deposited energy (see \cite{bolNIM} and references therein for a detailed description of this detector).

\begin{figure}
    \centering
    \includegraphics[width=0.3\textwidth]{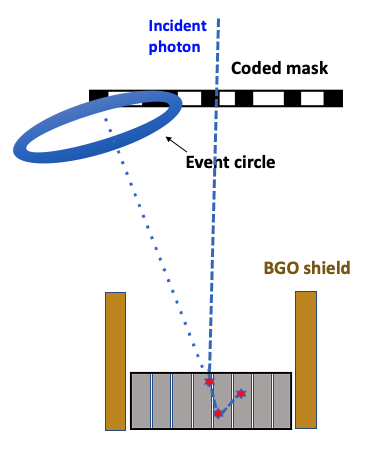}
    \caption{The CZT Imaging Calorimeter as a standalone Compton telescope and as a focal plane detector in a coded mask telescope. Grey rectangles represent the CZT bars, with red stars showing the detected photon interaction points. The dashed blue line shows the direction of the incident photon, while the dotted line shows the reconstructed direction of the Compton-scattered photon. The blue oval is the Compton-reconstructed event ring with its width reflecting the measurement accuracy.}
    \label{fig:cztsketch}
\end{figure}
The detected points of photon interactions in the CZT bars are used to reconstruct the event ring of the incident photons using the MEGAlib Compton analysis toolkit \cite{MEGAlib}, enabling the telescope to operate in Compton mode. The same analysis identifies the coordinates of the photon first interaction point, which along with its measured energy enables focal-plane detector capability for the coded aperture mask. 

The CsI calorimeter is positioned below the CZT Imaging calorimeter. It detects energy escaping from the CZT Calorimeter and measures the position of that energy deposition, improving the Compton reconstruction efficiency. All sides and the bottom of the CZT and CsI calorimeters are shielded by 4-cm thick BGO scintillator panels well, which efficiently absorbs natural and artificial background photons.

A coded aperture mask of GECCO is deployed at 20 meters above the CZT Imaging Calorimeter to increase the angular resolution, which is inversely proportional to the mask-detector separation. In this configuration the instrument aperture will be exposed to side-entering background radiation, which can significantly deteriorate the signal-to-noise ratio in coded mask imaging, and consequently the instrument sensitivity. This problem is solved by selecting events whose Compton-reconstructed direction points to the coded mask location. This is a unique feature of GECCO which greatly improves its angular resolution while maintaining a high signal-to-noise ratio.

The CZT Imaging Calorimeter, acting as a standalone Compton telescope with a large field-of-view, enables the coarse-scale measurement of ``total'' diffuse+point source emission, and also locates point sources with limited angular resolution. The coded-aperture mask provides the detection and localization of point sources, otherwise unresolved, with sub-arcminute angular resolution. Combining  the Compton telescope data with that obtained with the coded mask, GECCO will separate diffuse and point-source components in Galactic gamma-radiation with high sensitivity \cite{2015A&A...581A.126S}. An iterative analysis approach will enable GECCO to reveal faint sources and their characteristics as well as measuring actual diffuse radiation. 

GECCO can operate in either scanning or pointed mode. In scanning mode, it will observe the Galactic Plane. It will change to pointed mode to either increase observation time for special regions of interest, (e.g. the Galactic Centre) or to observe transient events such as flares of various origins or gamma-ray bursts. 
The expected GECCO performance is as follows \cite{Orlando:2021get}: energy resolution $<$ 1\% at 0.5--5 MeV, angular resolution $\sim$ 0.5 arcmin in mask mode with 5$^\circ$ field-of-view, and 4--8$^\circ$ in the Compton mode with $\sim$80$^\circ$ field-of-view. The effective area varies from 200 cm$^2$ to $\sim$ 2000 cm$^2$, depending on the energy.

\subsection{Instrument Sensitivity}

The major limiting factors to the instrument sensitivity are the backgrounds of different natures, and their efficient reduction and suppression are critical to any telescope in the MeV energy range. These backgrounds include bright albedo and Earth limb radiation, galactic diffuse radiation, background nuclear lines from the instrument and spacecraft, and nuclear lines produced by activation of the instrument and spacecraft by charged cosmic rays.
Both kinds of instrumental backgrounds have been carefully addressed in the INTEGRAL mission \cite{Lebrun, Sturner, Segreto, Weid2}, as well as in the preparations for ACT \cite{ACT}, COSI \cite{Zoglauer}, eASTROGAM \cite{Cumani, eASTROGAM} and capable simulation tools have been developed, e.g., MGGPOD \cite{Weid1} and MEGAlib \cite{MEGAlib}. 
The activation background is especially dangerous and very hard to counteract because this radiation usually is delayed after activation occurs and so cannot be simply eliminated by the anti-coincidence veto. Special attention has been paid to the CZT-created background \cite{Sim, Lim}.
Owing to that comprehensive studies, background suppression at the design level is implemented in GECCO by (but not limited to) the following:

 a)  to reduce the background from bright albedo and Earth limb radiation the GECCO detectors are placed inside  a thick active BGO shield, covering the sides and the bottom of the instrument. They absorb most  side- and bottom-entering gamma-radiation, both of primary (natural) and secondary origin, and also protect against dominating charged cosmic rays by creating a veto signal, 
 
b)  the equatorial low-Earth orbit (550-600 km altitude, $<5^\circ$ inclination) is chosen as optimal to minimize the effect of material activation by charged cosmic rays while crossing the South Atlantic Anomaly (SAA). For the same purpose the instrument design and material choice have been optimized: the mechanical structure has been designed with predominant use of composite (non- metal) materials,

c) a highly-efficient plastic-scintillator is placed on top of the CZT Imaging Calorimeter, vetoing $>$99.9\% of overwhelming flux of charged particles entering the detectors, 

d) the coded aperture mask is covered by a highly-efficient plastic scintillator which creates a veto signal to eliminate background secondary photons produced in the mask by incident charged cosmic rays

The determination of the future mission’s sensitivity is far from trivial: it always includes a number of critical assumptions. Some of the inputs to the sensitivity estimate are not well known, or not known at all in the early stages of the instrument's development. However, as the mission progresses, especially during orbital operation, the assessment of the sensitivity gradually increases due to better understanding of all the critical inputs, and especially due to continuously improving data analysis. Nevertheless, because GECCO’s sensitivity is a key parameter for the mission planning, and in particular for the content of this paper, we present here initial sensitivity estimates for GECCO.

The continuum (or point source) sensitivity can be estimated from the source detection confidence definition:
\begin{equation}
    n_\sigma=\frac{N_{\rm src}}{\sqrt{N_{\rm src}+B}}=
    \frac{I_{\rm src}\times A\times T\times \Delta E}{\sqrt{I_{\rm src} \times A\times T\times \Delta E+B}}
\end{equation}

\begin{figure}
    \centering
    \includegraphics[width=0.5\textwidth]{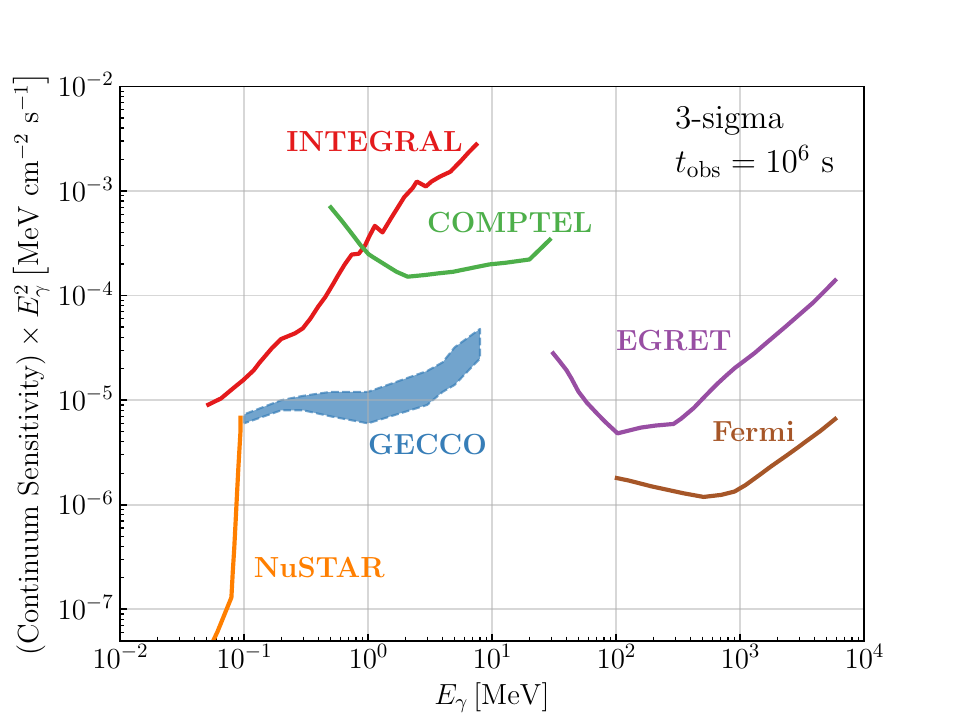}
    \caption{A (preliminary) comparison of instrumental sensitivities with GECCO's projected sensitivity, as calculated via \cref{eq:sens}), see the text for details.}
    \label{fig:sensitivities}
\end{figure}

From this equation, assuming $\Delta E =E$, the instrument sensitivity for a point-like source as seen within the instrumental point spread function of solid angle $\Delta\Omega$, as a function of the photon energy, can be derived as follows: 
\begin{equation}\label{eq:sens}
    S(E)=\frac{E}{2\times A_\mathrm{eff}(E) \times T}\left(n^2+n\times\sqrt{n^2+4\times B(E)}\right),
\end{equation}
where $B(E) = F_{\rm bckg} \times \Omega (E) \times T \times A_{\rm eff} (E)$ is the number of background counts, $E$ is the incident photon energy, n is a detection confidence level expressed in number of $\sigma$, $F_{\rm bckg}$ is the total background flux, $\Delta \Omega$ is a solid angle of the event acceptance, which in our case corresponds to the event circle (shown as blue ring in Fig. ~\ref{fig:cztsketch}), and $T$ is the observation time. We'd like to emphasize the importance of $\Delta \Omega$: if we did not use the Compton reconstruction to select the events for the analysis, it would be the full FoV of the telescope. The use of Compton reconstruction reduces it to the event circle and consequently reduces the background acceptance. The ``thickness'' of the event ring is defined by the instrument angular resolution, which is called Angular Resolution Measure (ARM) for the Compton reconstruction. The $\Delta\Omega$ is calculated as
\begin{equation}
    \Delta\Omega = 2 \pi \left[ \cos\left(c - \frac{d}{2}\right) - \cos(c + d) \right]
\end{equation}
where $c$ is the average Compton scattering angle, and $d$ is the ring width, equal to ARM/2. For this estimate we use the measured diffuse background $F_{bckg}$ from \cite{1999ApJ...520..124G}, and apply an additional ``safety'' factor of 3 to account for unknown contributions such as activation. The estimated GECCO sensitivity band we show in fig.~\ref{fig:sensitivities} is based on the most up to date currently simulated instrumental performance. The band size reflects the assumptions and uncertainties we use in our estimates (with further details offered in Ref.~\cite{Orlando:2021get}). The low-energy limit for the Compton measurements is about 200 keV due to rapidly decreasing Compton interaction cross-section yielding to photo-absorption, and for lower energy we instead use the mask-only, or ``classical'' coded mask analysis. For this analysis to create the mask image we need only the point of the first photon interaction in the focal plane detector, so we use single-site events which have only one interaction point in the detector, or use the first interaction point identified by the Compton reconstruction for multiple-hit events.  The effective background acceptance solid angle in this analysis is 0.85 sr, which is  the full GECCO FoV=1.5sr convolved with the $A_{\rm eff}(\Theta )$ but since the event statistics is rapidly increasing at lower energy, the sensitivity is rather good. The used lower energy limit of 100 keV is a conservative value of the CZT detector sensitivity, while the upper energy limit (10-15 MeV) is constrained by the CZT front-end electronics dynamic range, and also by our concept to stay in the Compton interaction energy range because the CZT drift-bar approach has poorer performance here due to the continuous energy deposition in the detector by charged particles (electron-positron components of the photon conversion at higher energy). 

With ongoing work on the improvement of the Compton event pattern reconstruction and background events recognition and removal (e.g. employing neural network techniques \cite{Zoglauer}), and the instrument optimization to reduce the activation background, it is feasible to noticeably improve the sensitivity subject to future project developments. Notice that this sensitivity analysis is strictly valid for a standalone source, or for a bright source surrounded by weaker neighboring sources. The detection of a faint source with a bright neighbor is a problem for the coded aperture mask technique, which is currently under investigation. Also, presently we are working on the combined full-size simulations of GECCO performance and sensitivity, to make more accurate sensitivity prediction.  The GECCO performance is particularly promising for searching for dark matter particles with O(MeV)-scale masses as well as for evaporating primordial black holes with $\mathcal{O}(10^{17}\, \mathrm{g})$ masses, as explained in the remainder of this work. Full details will be provided in the forthcoming published version of Ref.~\cite{Orlando:2021get}.

%%%%%%%%%%%%%%%%%%%%%%%%%%%%%%%%%%%%%%%%%%%%%%%%%%%%%%%%%%%

\section{Searches for Annihilating and Decaying Sub-GeV Dark Matter}
\label{sec:dm}

In this section we demonstrate that GECCO will be especially well-suited to search for particle dark matter (DM) in the MeV mass range. After reviewing DM indirect detection and explaining how we set limits using existing gamma-ray data and make projections for GECCO, we study the instrument's capabilities to detect the annihilation and decay of DM into specific Standard-Model final states. We also project GECCO's sensitivity reach for three specific, well-motivated DM models: one with an additional scalar mediating the DM's interaction with the Standard Model, a second one with a vector mediator and a third one in which the DM is an unstable right-handed neutrino. Throughout we utilize our code \texttt{hazma}, which we previously developed to analyze DM models producing MeV-scale gamma rays~\cite{hazma}.

\subsection{Indirect Detection Constraints and Projections}
\label{sec:id-procedure}

The prompt gamma-ray flux from DM annihilating or decaying in a region of the sky subtending a solid angle $\Delta\Omega$ is given by
\begin{widetext}
\begin{equation}
    \dv{\Phi}{E_\gamma}\Big|_{\bar{\chi}\chi}(E_\gamma) = \frac{1}{4\pi\, m_\chi^a} \cdot \left[ \int_{\Delta\Omega} \dd{\Omega} \int_{\mathrm{LOS}} \dd{l} [\rho(r(l, \psi))]^a \right] \cdot \Gamma \cdot \dv{N}{E_\gamma}\Big|_{\bar{\chi}\chi}(E_\gamma), \label{eq:flux}
\end{equation}
\end{widetext}
where ``LOS'' indicates the integral along the observation's direction line of sight. 
For decaying (annihilating) DM $a=1$ ($a=2$). The integral in the bracketed term ranges over lines of sight within a solid angle $\Delta\Omega$ from the  target region direction. This is referred to as the $D$ factor for decaying DM and $J$ factor for annihilating DM. It is proportional to the angle-averaged number of particles (pairs of particles) in the target available to decay (annihilate). The third term is the DM interaction rate. This is $\Gamma=1/\tau$ for decaying DM, where $\tau$ is the DM's lifetime. For annihilating DM, $\Gamma = \sv_{\bar{\chi}\chi} / 2 f_\chi$, where $f_\chi = 1$ if the DM is self-conjugate and $2$ otherwise (we assume the latter in this work). The last term is the photon spectrum per decay or annihilation. The calculation of this spectrum in \texttt{hazma} accounts for the radiative decay chains of the charged pion and muon as well as model-dependent final-state radiation from annihilations that produce electrons, muons and pions relevant for studying specific particle DM models.

To connect the gamma-ray flux with existing and future gamma-ray observations, we use a marginalized flux, given by
\begin{equation}
    {\dv{\bar{\Phi}}{E_\gamma}}\Big|_{\bar{\chi}\chi}(E_\gamma) \equiv \int \dd{E_\gamma'} R_\epsilon(E_\gamma | E_\gamma') {\dv{\Phi}{E_\gamma}}(E_\gamma').
\end{equation}
In the equation above, $R_\epsilon(E_\gamma | E_\gamma')$ is the telescope's energy resolution function, specifying the probability that a photon with true energy $E_\gamma'$ is detected with energy $E_\gamma$. This is well-approximated as a normal distribution $R_\epsilon(E_\gamma) = N(E_\gamma | E_\gamma', \epsilon E_\gamma')$~\cite{Bringmann:2008kj}, which defines $\epsilon$.\footnote{Note that the energy resolution of detectors is also sometimes given in terms of the full width at half maximum of this distribution.} 
To set an upper limit on the DM contribution to gamma-ray observations we perform a $\chi^2$ test with the quantity
\begin{equation}
    \chi_\mathrm{obs}^2 = \sum_i \qty( \frac{\operatorname{max} \qty [ \bar{\Phi}^{(i)}_{\bar{\chi}\chi} - \Phi^{(i)}_\mathrm{obs}, 0 ]}{\sigma^{(i)}} )^2\, , \label{eq:existing-constraints}
\end{equation}
where the sum ranges over energy bins, the flux in the numerator is the integral marginalized flux over bin $i$ and the denominator is the upper error bar on the observed integrated flux. Including an explicit background model would introduce significant systematic uncertainties since there is a paucity of MeV gamma-ray data, and in practice we expect it would only strengthen our constraints by less than an order of magnitude~\cite{Essig:2013goa}.\footnote{For final states containing monochromatic gamma rays the resulting constraints depend on the binning of the data. In the figures that follow we manually smooth out constraints in this case to account for different possible ways the data could have been binned.}

%================================
%---- NEW: Fisher ---------------
%================================

In order to estimated the discovery reach of GECCO, we apply the Fisher forecasting method developed in Ref.~\cite{Edwards:2017mnf} to account for imperfect knowledge of the background model. We choose as benchmark targets the Galactic Center as well as two nearby, extra-galactic targets: the Andromeda galaxy (M31), where tentative signals from dark matter decay in X-ray \cite{Jeltema:2014qfa} as well as in gamma-rays \cite{McDaniel:2018vam} have been claimed in recent years, and the Draco dwarf spheroidal galaxy, arguably one of the most promising among nearby, dark satellite galaxies with extremely low astrophysical gamma-ray background \cite{Colafrancesco:2006he, Jeltema:2015mee}.

We let the total differential flux from background and DM annihilations/decays be
\begin{align}
    \phi\qty(\bm{\theta}) =
    {\pdv{\Phi_{\chi}}{E_{\gamma}}{\Omega}}\qty(\bm{\theta}_{\chi}) +
    {{\pdv{\Phi_{\mathrm{bkg}}}{E_{\gamma}}{\Omega}}}\qty(\bm{\theta}_{\mathrm{bkg}}).
\end{align}
In the above expression, \(\bm{\theta}_{\chi}\) and \(\bm{\theta}_{\mathrm{bkg}}\) are the parameters of the DM and background differential fluxes and \(\bm{\theta}=\qty{\bm{\theta}_{\chi},\bm{\theta}_{\mathrm{bkg}}}\).

We parameterize the differential flux from DM with a single free parameter, \(\Gamma_{\chi}\), which specifies the normalization. In the case of DM annihilations, \(\Gamma_{\chi}\) is taken to be the velocity-averaged annihilation cross section \(\expval{\sigma_{\bar{\chi}\chi} v}\), while for DM decays, \(\Gamma_{\chi}\) is the inverse DM lifetime \(1/\tau\). To model the background from the Galactic Center, we follow Ref.~\cite{ray2021near}, including a galactic contribution adapted from Ref.~\cite{Bartels:2017dpb} and an extra-galactic contribution. The galactic contribution consists of several spectral templates computed with \texttt{GALPROP}\footnote{\href{http://galprop.stanford.edu}{http://galprop.stanford.edu}}~\cite{Strong:2001fu} and an analytic component, tailored to fit existing gamma-ray data in the inner part of the Milky Way. 

We note that a possible point-source contribution contamination from Sag A$^*$ associated with 4FGL J1745.6$-$2859 is not excluded, but recent studies show that it would significantly dimmer than the extended emission we consider in searching for dark matter in the Galactic Center region (see e.g. \cite{Cafardo:2021pqs} and references therein, and in particular their estimate of the source emission from 4FGL J1745.6$-$2859 in the 60-300 MeV range).

Our full background model for the Galactic Center contains six parameters and is given by:
\begin{widetext}
\begin{align}
    \pdv{\Phi_{\mathrm{GC}}}{E_{\gamma}}{\Omega} = 
    A_{\mathrm{g}}\qty(\frac{E_{\gamma}}{1 \ \mathrm{MeV}})^{-\alpha_{\mathrm{g}}}
    \exp(-\qty(\frac{E_{\gamma}}{E_c})^{\gamma}) 
    + A_{\mathrm{e.g.}}\qty(\frac{E_{\gamma}}{1 \ \mathrm{MeV}})^{-\alpha_{\mathrm{e.g.}}}.
\end{align}
\end{widetext}
where \(A_{\mathrm{g}/\mathrm{e.g.}}\) are the amplitudes, \(\alpha_{\mathrm{g}/\mathrm{e.g.}}\) the power-law indices, \(E_{c}\) the exponential cut-off and \(\gamma\) the exponential index.
The subscripts ``g'' and ``e.g.'' stand for ``galactic'' and ``extra-galactic''. 
The galactic component has the same form as the ``$\mathrm{ICS}_\mathrm{lo}$'' component from Ref.~\cite{Bartels:2017dpb}. This is the dominant background over GECCO's energy range, and required to fit COMPTEL data in the galactic center. We use the same fiducial parameter values for the normalization, power-law index and cutoff energy as Ref.~\cite{Bartels:2017dpb}. The extragalactic term, for which we use the fiducial values from Ref.~\cite{ray2021near}, dominates below $\sim 0.3\, \mathrm{MeV}$.

For observations in the directions of Draco and M31, we use a simpler power law simultaneously accounting for the galactic and extragalactic background:
\begin{align}
    \pdv{\Phi_{\mathrm{EG}}}{E_{\gamma}}{\Omega} = 
    \bar{A}\qty(\frac{E_{\gamma}}{1 \ \mathrm{MeV}})^{-\bar{\alpha}}.
\end{align}
As the fiducial parameter values we use the fit to high-latitude COMPTEL and EGRET data from Ref.~\cite{Boddy:2015efa}. We note that M31 has been detected in gamma rays by the Fermi telescope \cite{Fermi-LAT:2010zba}. However, it is nontrivial to extrapolate the faint detected emission to the lower energies relevant here.

Given observations at higher gamma-ray frequencies, where M31 is detected at the $5 \sigma$ level \cite{Fermi-LAT:2010zba}, it is to be expected that some astrophysical background exist from M31 as well; discrimination of a dark matter signal from such background will entail the use of spectral as well as morphological information, and possibly multiwavelength observations, along the lines of e.g. what discussed in Ref.~\cite{McDaniel:2018vam}.

\begin{table*}\centering
    \begin{tabular}{p{1.5cm} c l c}
        \toprule
        Target & Parameter & Description & Fiducial value \\
        %-----
        \midrule
        \multirow{6}{*}{\rotatebox[origin=c]{90}{Galactic Center}} & \(A_{\mathrm{g}}\) &  Galactic amplitude & \(0.013 \ \qty[\mathrm{MeV}^{-1}\mathrm{cm}^{-2}\mathrm{s}^{-1}\mathrm{sr}^{-1}]\) \\
        %-----
        & \(\alpha_{\mathrm{g}}\) &  Galactic power-law index & \(1.8\) \\
        %-----
        & \(E_{c}\) &  Exponential cutoff energy & \(2 \ \qty[\mathrm{MeV}]\) \\
        %-----
        & \(\gamma\) &  Exponential cutoff index & \(2\) \\
        %-----
        & \(A_{\mathrm{e.g.}}\) &  EG amplitude & \(0.004135 \ \qty[\mathrm{MeV}^{-1}\mathrm{cm}^{-2}\mathrm{s}^{-1}\mathrm{sr}^{-1}]\) \\
        %-----
        & \(\alpha_{\mathrm{e.g.}}\) &  EG power-law index & \(2.8956\) \\
        %-----
        \midrule
        %-----
        \multirow{2}{*}{\rotatebox[origin=c]{90}{\parbox{1.5cm}{M31 \& Draco}}} & \(\bar{A}\) & Amplitude & \(2.4\times 10^{-3} \ \qty[\mathrm{MeV}^{-1}\mathrm{cm}^{-2}\mathrm{s}^{-1}\mathrm{sr}^{-1}]\) \\
        %-----
        & \(\bar{\alpha}\) & Power-law index & \(2\) \\
        %-----
        \bottomrule
    \end{tabular}
    \caption{Fiducial values of the background model parameters.}
    \label{tab:bkg_params}
\end{table*}

To compute the upper limit on the DM annihilation/decay rate \(\Gamma_{\chi}\), we start by computing the Fisher matrix~\cite{Edwards:2017mnf}
\begin{align}
    \mathcal{F}_{ij} 
    &=
    \int
    \dd{E_{\gamma}}
    \dd{\Omega}
    T_{\mathrm{obs}}
    A_{\mathrm{eff}}
    \eval{\qty(
    \frac{1}{\phi}
    \pdv{\phi}{\theta_{i}}
    \pdv{\phi}{\theta_{j}}
    )}_{\bm{\theta}=\bm{\theta}_{\mathrm{fid.}}}
\end{align}
where we chose \(\theta_{1} = \Gamma_{\chi}\) and \(\bm{\theta}_{\mathrm{fid.}}\) are the fiducial values of the parameters with \(\Gamma_{\chi}\) set to zero. 
Our Fisher matrix is a \(7\times7\) symmetric matrix for observations of the GC and \(3\times3\) for M31 and Draco. Lastly, the estimated upper limit on the DM annihilation/decay rate is computed using~\cite{Edwards:2017mnf}
\begin{align}
    \Gamma_{\chi}^{\mathrm{UL}} &= N_{\sigma}\sqrt{\qty(\mathcal{F}^{-1})_{11}}
\end{align}
For all of our limits, we take \(N_{\sigma} = 5\) as the detection threshold.

The $J$ and $D$ factors for the GECCO targets are shown in \cref{tab:J_gecco}.\footnote{
    Note that the profile we use for Draco gives $J$ and $D$ factors a factor of $\sim 2$ larger than more recent works that use NFW~\cite{2019MNRAS.482.3480P} and more general density profiles~\cite{Hayashi:2020jze} for a $0.5^\circ$ observing region. This difference is within about $2\sigma$ of the uncertainties on the $J$ and $D$ factors' values.
} These are derived from fits of dark matter density profiles to measurements of the targets rotation curves, surface brightnesses and velocity dispersions. We employ a Navarro-Frenk-White (NFW) density profile~\cite{Navarro:1995iw} for all targets and additionally consider an Einasto profile~\cite{EinastoProfile} for the Galactic Center to bracket the uncertainties in our analysis stemming from assumptions about the dark matter distribution, with references given in the table. For our analysis of annihilating DM we select a $1'$ observing region (roughly GECCO's angular resolution) to maximize the signal-to-noise ratio. In the case of decaying DM we instead find the best strategy is to use a larger $5^\circ$ field of view, since the $D$ factor depends much less strongly on the observing region's size. The observing regions and the $J$ and $D$ factors used to collect existing gamma-ray data are presented in \cref{tab:J_existing}. We note that one could possibly consider dark matter annihilation or decay at all redshifts \cite{Ajello:2015mfa}; the predicted signal strength is generally predicted to be weaker than from the targets we consider, and prone to significant uncertainties due to the largely unknown clustering properties of dark matter halos as a function of redshift. 

\emph{Secondary} photons are also produced by dark matter processes that create electrons and positrons. These can produce energetic photons via inverse-Compton scattering against ambient CMB, starlight and dust-reprocessed infrared photons~\cite{CPU1, CPU2}. Their spectrum, for upscattered initial photon energy $E_\gamma$ peaks near $E_{\rm peak}\simeq E_\gamma(E_e/m_e)^2\simeq E_\gamma (m_{\rm DM}/(10\, m_e))^2$ which for sub-GeV DM masses and for the highest energy background photon from starlight ($E_\gamma\sim 1$ eV) gives $\lesssim 100~\mathrm{keV}$ upscattered photon energy, thus well below GECCO's expected energy threshold. Also, the calculation of the secondary radiation carries inherently difficult systematics ranging from the effects of diffusion to the morphology of the background radiation fields.

Observations of the cosmic microwave background (CMB) constrain the amount of power DM annihilations and decays are allowed to inject in the form of ionizing particles during recombination~\cite{Chen:2003gz,Padmanabhan:2005es,Galli:2009zc,Slatyer:2015jla,Aghanim:2018eyx}. \texttt{hazma} contains functions for calculating this constraint for annihilating DM. To review, given a DM model the constraint is set by
\begin{equation}
    p_\mathrm{ann} = f_\mathrm{eff}^\chi \frac{\sv_{\bar{\chi}\chi,\mathrm{CMB}}}{m_\chi}, \label{eq:cmb}
\end{equation}
where $f_\mathrm{eff}^\chi$ is the fraction of energy per DM annihilation imparted to the plasma and $p_\mathrm{ann}$ is an effective parameter measured from observations bounding the energy that can be injected per unit volume and time. In turn $f_\mathrm{eff}^\chi$ depends on the photon and electron/positron spectrum per DM annihilation.

If the DM self-annihilation cross section is $s$-wave (i.e. velocity-independent), the quantity $\sv_{\bar{\chi}\chi,\mathrm{CMB}}$ is equal to the present-day self-annihilation cross section. If instead the DM annihilates in a $p$-wave (i.e. is velocity-suppressed), the present-day self-annihilation cross section is related to the one at CMB via the squared ratio of the DM velocity at present and at recombination, $(v_{\chi,0} / v_{\chi,\mathrm{CMB}})^2$. Computing $v_{\chi,\mathrm{CMB}}$ requires the DM's kinetic decoupling temperature as input, which is model-dependent.

The kinetic decoupling temperature is the point at which momentum transfer between the thermal bath and the DM becomes slow compared to the Hubble rate. More quantitatively, the rate of momentum transfer is roughly the product of the density of the SM bath, the elastic DM-SM scattering cross section and the number of scatterings required to substantially alter a DM particles' momentum~\cite{Schmid:1998mx}:
\begin{equation}
    \Gamma_\mathrm{transfer} \sim n_\mathrm{SM} \, \sigma_{\mathrm{DM+SM}\to\mathrm{DM+SM}} \, \left( \frac{\delta p}{p} \right)^2 \, .
\end{equation}
Here $p \sim \sqrt{m_\chi T}$ is the momentum of a DM particle and $\delta p \sim T$ is the momentum change during a collision. The required scattering cross section is a model-dependent quantity. Considering the Higgs portal model we will study in \cref{sec:higgs_portal} as an example, the cross section for scattering elastically with electrons is approximately $(g_{S\chi} \, \sin\theta \, y_e)^2 / m_\chi^2$ by dimensional analysis, where $y_e$ is the electron Yukawa coupling. Since the density of the SM bath scales as $T^3$, equating the momentum transfer rate with the Hubble rate yields
\begin{align}
    \Gamma_\mathrm{transfer} \sim H \implies T^3 \, \frac{(g_{S\chi} \sin\theta y_e)^2}{m_\chi^2 v_H^2} \, \frac{T}{m_\chi} \sim \frac{T^2}{M_\mathrm{Planck}} \, .
\end{align}
Solving this gives an estimate of the kinetic decoupling temperature:
\begin{equation}
    T_\mathrm{kd} \sim \frac{1}{g_{S\chi} \sin\theta y_e} \sqrt{\frac{m_\chi}{M_\mathrm{Planck}}} \, .
\end{equation}
For values of $g_{S\chi} \sin\theta y_e$ consistent with existing experimental probes (see e.g. fig.~2 of Ref~\cite{Krnjaic:2015mbs}), $T_\mathrm{kd} \gtrsim 10^{-6}$. This is also in line with assumptions from previous MeV-scale DM studies~\cite{Essig:2013goa}. Therefore in the following sections we fix $T_\mathrm{kd} = 10^{-6}$ when demonstrating CMB limits on $p$-wave annihilating DM, and comment on how the bound would vary for higher values.

For constraints on decaying DM we reuse the CMB limits derived in Ref.~\cite{Slatyer:2016qyl}.

\begin{table}[]
    \centering
    \begin{tabular}{ccccc}
        \toprule
        Target & $J(1')$ & $J(5^\circ)$ & $D(1')$ & $D(5^\circ)$  \\
        \midrule
        Galactic Center (NFW)~\cite{deSalas:2019pee} & $1.853\ee{26}$ & $4.259\ee{28}$ & $1.286\ee{20}$ & $3.817\ee{24}$\\
        Galactic Center (Einasto)~\cite{deSalas:2019pee} & $1.591\ee{28}$ & $1.187\ee{30}$ & $1.111\ee{21}$ & $4.919\ee{24}$\\
        Draco (NFW)~\cite{Dugger:2010ys} & $9.085\ee{23}$ & $1.926\ee{25}$ & $1.581\ee{19}$ & $4.747\ee{22}$\\
        M31 (NFW)~\cite{2015PASJ...67...75S} & $3.976\ee{24}$ & $3.535\ee{25}$ & $8.763\ee{19}$ & $9.601\ee{22}$\\
        \bottomrule
    \end{tabular}
    \caption{{$J$ and $D$ factors for various circular targets, in units of $\mathrm{MeV}^2\, \mathrm{cm}^{-5}$ and $\mathrm{MeV}\, \mathrm{cm}^{-2}$ respectively.} The dark matter profile parameters are taken from the indicated references. For the Milky Way, we use the values from Table III of Ref.~\cite{deSalas:2019pee}. The Einasto profile parameters are adjusted within their $1\sigma$ uncertainty bands to maximize the $J$ and $D$ factors. For all other targets we use the parameters' central values. The distance from Earth to the Galactic Center is set to 8.12 kpc~\cite{Abuter:2018drb,deSalas:2019pee}. For reference, the angular extents of the $1'$ and $5^\circ$ regions are $2.658\ee{-7}\, \mathrm{sr}$ and $2.39\ee{-2}\, \mathrm{sr}$ respectively.}
    \label{tab:J_gecco}
\end{table}

\begin{table}[]
    \centering
    \begin{tabular}{ccccc}
        \toprule
        Experiment & Region & $\Delta\Omega$ [sr] & $J$ & $D$\\
        \midrule
        COMPTEL~\cite{Kappadath:1993} & $|b| < 20^\circ,\, |l| < 60^\circ$ & 1.433 & $1.333\ee{29}$ & $5.973\ee{25}$\\
        EGRET~\cite{Strong_2004} & $20^\circ < |b| < 60^\circ,\, |l| < 180^\circ$ & 6.585 & $4.126\ee{28}$ & $1.126\ee{26}$\\  % UPDATED ON 23 MARCH
        Fermi~\cite{Ackermann:2012pya} & $8^\circ < |b| < 90^\circ,\, |l| < 180^\circ$ & 10.82 & $9.170\ee{28}$ & $1.928\ee{26}$\\  % UPDATED ON 23 MARCH
        INTEGRAL~\cite{Bouchet:2011fn} & $|b| < 15^\circ,\, |l| < 30^\circ$ & 0.5421 & $1.131\ee{29}$ & $3.957\ee{25}$\\
        \bottomrule
    \end{tabular}
    \caption{{$J$ and $D$ factors for observing regions in the Milky Way used by past experiments, in units of $\mathrm{MeV}^2\, \mathrm{cm}^{-5}$ and $\mathrm{MeV}\, \mathrm{cm}^{-2}$ respectively.} The regions are specified in Galactic coordinates. We again use the NFW profile parameters from Table III of ref.~\cite{deSalas:2019pee}.}
    \label{tab:J_existing}
\end{table}

\subsection{Model-independent projections}

\begin{figure*}
    \centering
    \includegraphics[width=\textwidth]{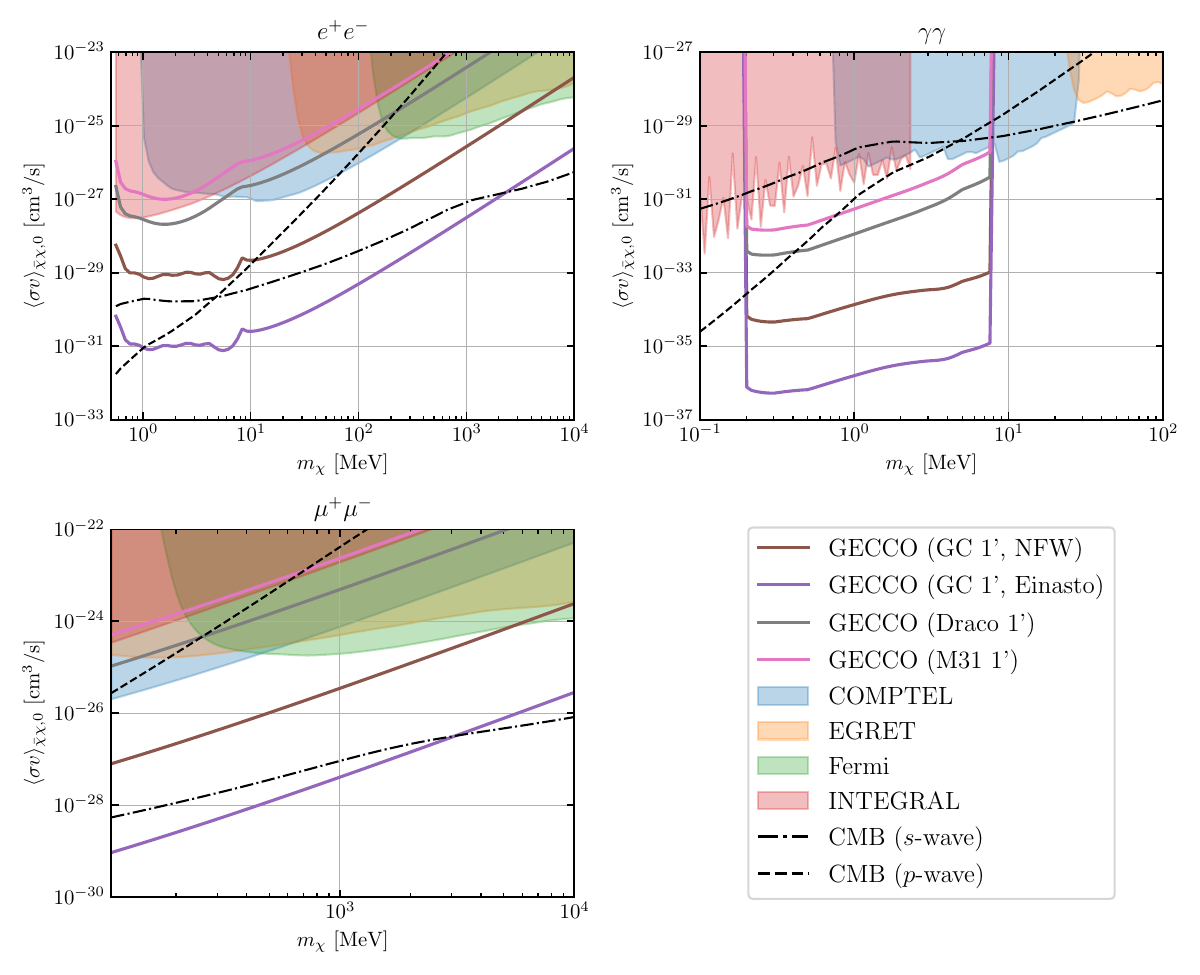}
    \caption{Projected constraints on annihilation into different final states (solid lines). 
    The shaded regions show constraints from existing gamma ray data. The dashed black line shows the CMB constraint assuming the DM annihilation are $p$-wave and have a kinetic decoupling temperature of $10^{-6}\,  m_\chi$; higher kinetic decoupling temperatures would give weaker constraints. The dot-dashed line gives the CMB constraint for $s$-wave DM annihilations.}
    \label{fig:single_channel_ann_limits}
\end{figure*}

\begin{figure*}
    \centering
    \includegraphics[width=\textwidth]{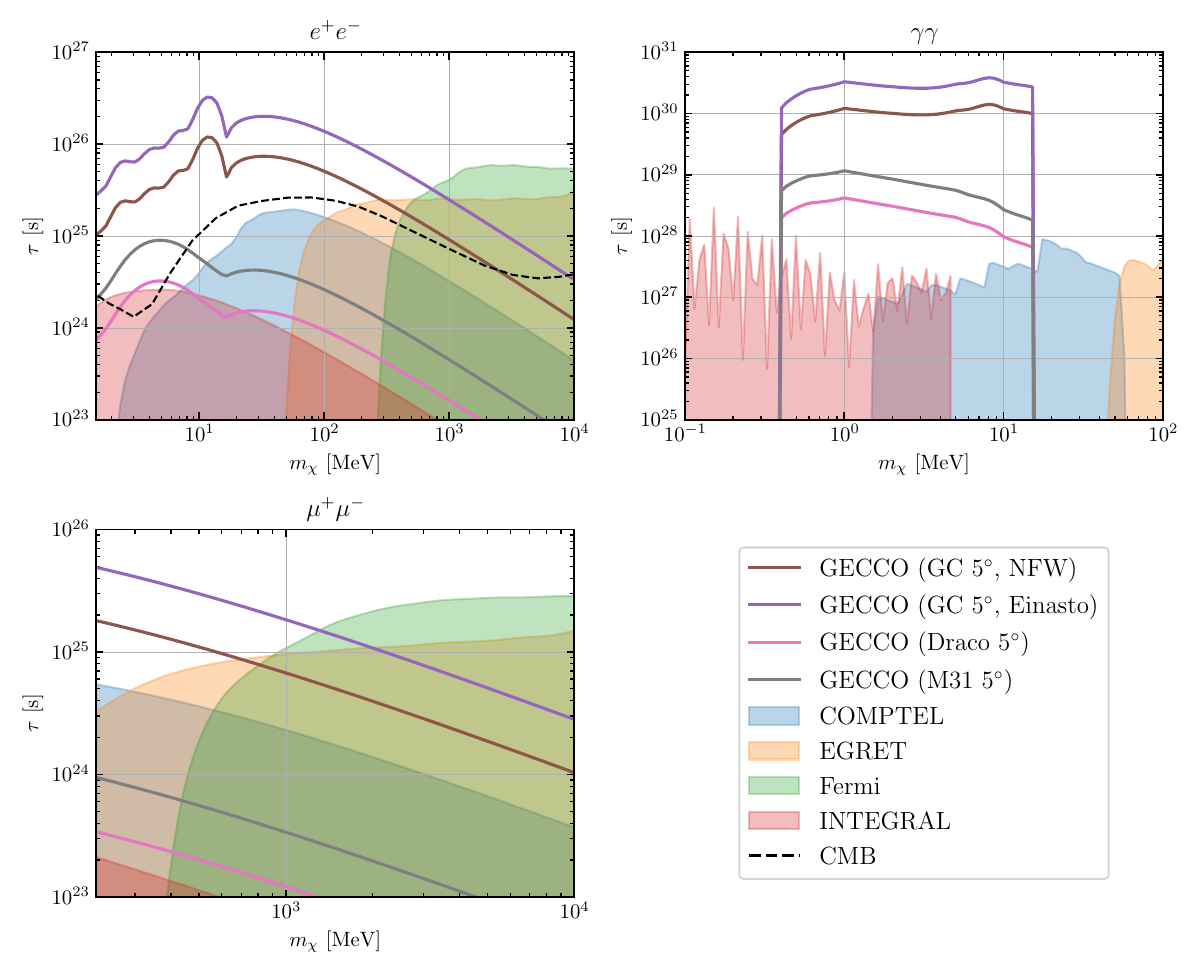}
    \caption{Constraints on the DM particle's lifetime for decays into different final states (solid lines). To account for the unknown systematics of GECCO, the surrounding bands show how the projections would change if the background photon counts were a factor of 25 higher than the fiducial value. The CMB constraint on decays into $e^+ e^-$ is taken from Ref.~\cite{Slatyer:2016qyl}. While constraints for the $\mu^+ \mu^-$ final state are not provided, we estimate they lie around $10^{24} - 10^{25}\, \mathrm{s}$ since the subsequent muon decays produces electrons with energy $\sim 1/3 m_\chi$. The constraint for decays into $\gamma \gamma$ lies below the axis range.}
    \label{fig:single_channel_lifetime_limits}
\end{figure*}

We first consider GECCO's discovery reach for ``simplified'' dark matter models where the dark matter particles annihilate or decay into exclusive, single final states, namely the diphoton, dielectron and dimuon final states.\footnote{
    The results for annihilation into two pions are weaker than the results for the dimuon final state by an order one factor, but otherwise nearly identical, so we do not plot them separately.
} The existing gamma-ray constraints and GECCO projections on the branching fraction times self-annihilation cross section (for annihilating DM) are shown in \cref{fig:single_channel_ann_limits} and on the lifetime (for decaying DM) in \cref{fig:single_channel_lifetime_limits}. In the figures we shade regions of parameter space ruled out by observations taken with previous or existing telescopes according to our limit-setting procedure described near \cref{eq:existing-constraints}. Our limits are based on data from COMPTEL~\cite{Kappadath:1993}, EGRET~\cite{Strong_2004}, Fermi-LAT~\cite{Ackermann:2012pya}, and INTEGRAL~\cite{Bouchet:2011fn}, and provide details on the regions of interest and $J$ and $D$ factors in \cref{tab:J_existing}. We also indicate constraints from CMB distortions with dashed and dot-dashed black lines (the regions excluded are above those lines).

There are relatively few analyses of existing gamma-ray data that overlap with the mass range we focus on. For comparison, our limits from existing data are close to those from Refs.~\cite{Boddy:2015efa} and \cite{Essig:2013goa} since they were set with a similar procedure. The analysis of 11 years of Fermi observations of 27 dwarf spheroidals in Ref.~\cite{Hoof:2018hyn} found limits on the self-annihilation cross section $10 - 100$ times stronger than ours for the $e^+ e^-$ and $\mu^+ \mu^-$ channels. This scaling can be accounted for by their substantially longer observing time ($3.5\times 10^{8}\, \mathrm{s}$ versus our $10^6\, \mathrm{s}$), their use of stacking and their careful background modeling. On the other hand, their constraints only extend down to $2\,\mathrm{GeV}$. Ref.~\cite{Cirelli:2020bpc} recently studied constraints from INTEGRAL on \emph{secondary} photons produced by MeV-scale DM and found stronger constraints than our for the $e^+ e^-$ and $\mu^+ \mu^-$ annihilation channels over the plotted mass range ($\sv_{\bar{\chi}\chi,0} \lesssim 10^{-27} - 10^{-25}\, \mathrm{cm}^3/\mathrm{s}$). However, uncertainties in the astrophysics of secondary emission can relax their bounds by an order of magnitude, bringing them into line with constraints on primary emission obtained using other telescopes.

The GECCO sensitivity is shown for four distinct cases, listed here from top to bottom in the order the lines appear in \cref{fig:single_channel_ann_limits} (the order is inverted for the lifetime in the case of decay shown in \cref{fig:single_channel_lifetime_limits}): 
the blue line corresponds to observations, within an angular region of $1^\prime$, of the Draco dSph; 
the magenta line for observations of M31, within the same angular region of $1^\prime$; 
finally the red and yellow lines correspond to observations of the Galactic Center, again within $1^\prime$, assuming an NFW profile (yellow line) and an Einasto profile (red line). 

We find that the greatest gains a telescope such as GECCO will bring in the search for MeV dark matter are for final states producing monochromatic gamma ray (i.e. lines). In this case the improvements to the sensitivity across the range between 0.1 and 10 MeV are forecast to be as large as four orders of magnitude in the annihilation rate, or over two orders of magnitude in lifetime. Signals will potentially be visible across different targets. The complementarity with CMB constraints depends on whether the DM annihilation is $s$-wave or $p$-wave (and its kinetic decoupling temperature in the $p$-wave case). This is not uniquely specified given just the DM self-annihilation cross section. The entire parameter space testable with GECCO is compatible with constraints from CMB for $p$-wave DM annihilations under the assumption the kinetic decoupling temperature is higher than $10^{-6} \, m_\chi$. GECCO observations have the potential to discover DM annihilating in an $s$-wave to two photons. While the $s$-wave CMB bounds for the dielectron and dimuon final states are more stringent, GECCO still has the potential to uncover DM annihilation in the Galactic Center depending on the DM mass and spatial distribution. 

The electron-positron final state also offers highly promising prospects, especially at low masses around 1-10 MeV, with improvements to the current sensitivity of up to 4 orders of magnitude in annihilation rate (two in lifetime) but will improve by an order of magnitude even at large masses, around 10 GeV; detection of an annihilation signal outside the Milky Way center will be possible again, but only for masses below an MeV or so, with similar prospects for decay.

Finally, in the muon pair case, the optimal dark matter candidate would have a mass of around the muon mass, offering an improvement of three orders of magnitude for annihilation, and over one in decay. However, in the $\mu^+\mu^-$ case current constraints exclude the possibility of detecting a signal from M31 or Draco, in either annihilation or decay.

In what follows we illustrate with explicit model realizations the physics reach of GECCO for the detection of dark matter annihilation in the Higgs portal (\cref{sec:higgs_portal}) and vector portal/dark photon (\cref{sec:vector}) cases, and of dark matter decay in the case the right-handed neutrino dark matter (\cref{sec:rhn}).

\subsection{Model Example: Higgs Portal}
\label{sec:higgs_portal}

In this model, we extend the Standard Model by adding a new scalar singlet \(\tilde{S}\). The dark matter interacts only with this scalar, through a Yukawa interaction: \(\mathcal{L}\supset g_{S\chi}\tilde{S}\bar{\chi}\chi\). The new scalar mixes with the real neutral scalar component of the Higgs with a mixing angle \(\theta\) providing a portal through which the dark matter can interact with the Standard Model.\footnote{
    This is achieved by modifying the scalar potential to be:
    \begin{align}
        V(\tilde{S}, H) = -\mu_H^2 H^{\dagger}H + \lambda \qty(H^{\dagger}H)^2 + \frac{1}{2}\mu_{S}^2\tilde{S}^2 + g_{SH}\tilde{S}H^{\dagger}H + \cdots
    \end{align}
    where \(H\) is the SM Higgs doublet, \(\tilde{S}\) is a new, neutral scalar singlet and the \(\cdots\) represent interaction terms with more than a single \(\tilde{S}\). After diagonalizing the scalar mass matrix we find two neutral scalars \(h\) and \(S\) which are related to the original scalars through a mixing angle: $\tilde{S} = h\sin\theta + S\cos\theta$ and $\tilde{h} = h\cos\theta - S\sin\theta$.
} This results in a Lagrangian density of the form:
\begin{widetext}
\begin{align}
    \mathcal{L} &=&  \mathcal{L}_{\mathrm{SM}} + \bar{\chi}\qty(i\cancel{\partial}-m_{\chi})\chi -\frac{1}{2}S\qty(\partial^2+m_{S}^2)S- g_{S\chi}(h\sin\theta + S\cos\theta) \bar{\chi}\chi + (h\cos\theta - S\sin\theta)\sum_{f}m_{f}\bar{f}f + \cdots 
    \end{align}
\end{widetext}
where \(f\) is a massive SM fermion and the \(\cdots\) contain pure scalar interactions. This Lagrangian density is only valid for energies \(E\gtrsim \Lambda_{\mathrm{EW}}\) while our interest lies in sub-GeV energies. To obtain a Lagrangian valid for sub-GeV energies, we first need to find a Lagrangian valid above the QCD confinement scale and then match onto the chiral Lagrangian  (see Ref.~\cite{scherer2005chiral} for a detailed review of chiral perturbation theory). We omit the details here (to be provided in a forthcoming paper) and simply give the result:
\begin{align}
    \label{eq:LagIntS}
    \mathcal{L}_{\mathrm{Int}(S)} & = 
    \frac{2 \sin\theta}{3 v_{h}} S 
    \left[ (\partial_\mu \pi^0) (\partial^\mu \pi^0) + 2 (\partial_\mu \pi^+) (\partial^\mu \pi^-) \right] \\
    & \hspace{1cm} + \frac{4 i e \sin\theta}{3 v_{h}} S A^\mu \left[ \pi^- (\partial_\mu \pi^+) - \pi^+ (\partial_\mu \pi^-) \right]\notag\\
    & \hspace{1cm} -\frac{m^2_{\pi^{\pm}}\sin\theta}{3v_{h}}\qty(\frac{5}{2}S + \frac{\sin\theta}{3v_{h}}S^2)\left[ (\pi^0)^2 + 2 \pi^+ \pi^- \right]\notag\\
    & \hspace{1cm} - \frac{10 e^2 \sin\theta}{27v_{h}} S \pi^+ \pi^- A_\mu A^\mu\notag \\
    & \hspace{1cm} - g_{S \chi} S \bar{\chi} \chi - \sin\theta S \sum_{\ell=e,\mu} \frac{y_\ell}{\sqrt{2}} \bar{\ell} \ell.\notag
\end{align}
In the equation above, we have made the redefinition $g_{S\chi} \cos\theta \to g_{S\chi}$. The terms relevant for indirect detection are those involving an $S$ field interacting with pions (along with a photon), leptons or dark matter. The $S^2 \pi \pi$ and $S \pi \pi A A$ terms are subdominant since they have additional factors of $\sin\theta$, the Higgs vev and/or the electron charge.

As discussed in our previous work~\cite{hazma}, this leading-order chiral perturbation theory approach has a limited regime of validity. To avoid the $f_0(500)$ resonance~\cite{f0500} and the resulting final-state interactions between pairs of pions as well as $(500\, \mathrm{MeV} / \Lambda_\mathrm{QCD})^2 \sim 20\%$ corrections from the next-to-leading order chiral Lagrangian~\cite{Pich:1995bw}, we restrict $m_\chi < 250\, \mathrm{MeV}$ when the DM annihilates into SM particles, and $m_S < 500\, \mathrm{MeV}$ when it predominantly annihilates into mediators.

The thermally-averaged DM self-annihilation cross section for this model is $p$-wave suppressed: $\sv_{\bar{\chi}\chi} \propto T_\chi / m$ for low DM temperatures $T_\chi$. Since this assumption holds for all our targets, under the assumption that the DM particles' speeds follow a Maxwell-Boltzmann distribution we can approximate $\sv_{\bar{\chi}\chi} \propto \sigma_v^2$, where $\sigma_v$ is the velocity dispersion in the target. We take $\sigma_v = 10^{-3}\, c$ for the Milky Way targets~\cite{2010AJ....139...59B} and M31~\cite{1980ApJ...242...53W} and $\sigma_v = 3\ee{-5}\, c$ for Draco~\cite{2020A&A...633A..36M}.\footnote{
    A more careful treatment would average over the position-dependent velocity distribution in the target. In the case of the Milky Way, this should only change our results by a factor of $\lesssim 2$~\cite{Boddy:2015efa}.
}

\begin{figure}
    \centering
    \includegraphics[width=\textwidth]{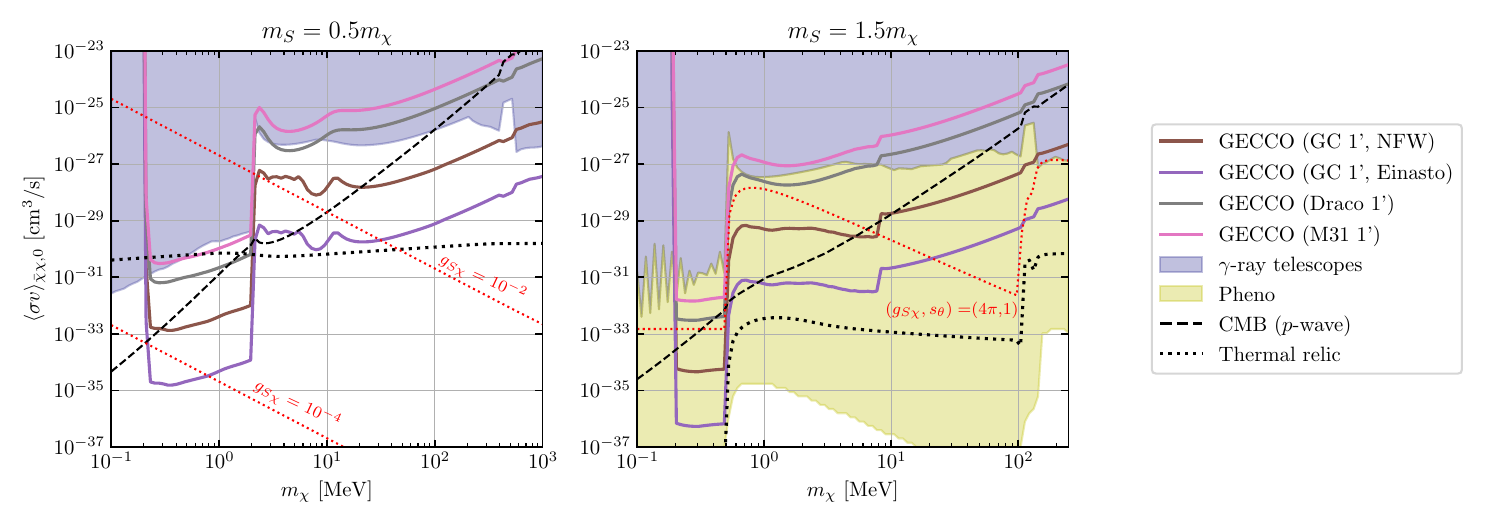}
    \caption{Constraints on the thermally-averaged DM self-annihilation cross section in the Milky Way for the Higgs portal model (solid lines). 
    The case where the indirect detection signal comes from annihilations into mediators (SM particles) is shown on the left (right). The thin red dotted lines are contours of constant coupling strength. The orange region in the right panel is a conservative exclusion region from experiments besides gamma-ray telescopes. The CMB constraint was computed assuming a kinetic decoupling temperature of \(10^{-6}  m_\chi\).}
    \label{fig:higgs_portal}
\end{figure}

The constraints from current gamma-ray data, our projections for GECCO's reach using different targets and the CMB bounds for this model are displayed in \cref{fig:higgs_portal}, with two ratios of $m_S$ to $m_\chi$. We have rescaled the constraints on $\sv_{\bar{\chi}\chi}$ for each target into constraints on $\sv_{\bar{\chi}\chi,0}$, the thermally-averaged self-annihilation cross section in the Milky Way. 
An array of terrestrial, astrophysical and cosmological observations constrain this Higgs portal model (see e.g. Ref.~\cite{Krnjaic:2015mbs}). Depending on the DM and mediator masses the most relevant ones for this work include rare and invisible decays of $B$ and $K$ mesons and beam dumps sensitive to visible $S$ decays into leptons. How these complement indirect detection bounds depends strongly on whether the DM annihilates into mediator pairs ($m_\chi > m_S$, left panel) or SM particles ($m_\chi < m_S$, right panel). In the first case, the DM self-annihilation cross section scales as $\sv_{\bar{\chi}\chi,0} \sim g_{S\chi}^4$, while other probes (including CMB energy injection constraints) bound $\sin\theta$. This means that as long as \emph{some} value of $\sin\theta$ is allowed, these probes do not constrain the strength of possible gamma-ray signals. This is indeed the case: while e.g. beam dumps and CMB observations bound $\sin\theta$ from above, there is a substantial gap between the lower bound on $\sin\theta$ from the requirement that decays of $S$ do not disrupt the predictions of big bang nucleosynthesis (BBN). Moreover, the BBN constraints are dependent on the assumption the universe had a standard thermal history. Without that assumption, $\sin\theta$ can be taken to be arbitrarily small. Since there are thus no constraints to plot (aside from those from existing gamma-ray telescopes), in the left panel of the \cref{fig:higgs_portal} we instead show contours of constant $g_{S\chi}$ to give a sense of reasonable values of the cross section. GECCO observations of the Galactic Center will probe down to $g_{S\chi} \sim 5\ee{-5}$ for low DM masses.

When the $S S$ final state is not accessible, the DM's annihilations are strongly suppressed since the cross section scales as $\sv_{\bar{\chi}\chi,0} \sim g_{S\chi}^2\, \sin^2\theta\, y^2$, where $y \ll 1$ is the Yukawa for the heaviest-accessible final state. This means correspondingly large values of the couplings are required to give indirect detection signals. The red line in the right panel of the figure shows the DM self-annihilation cross section for $(g_{S\chi},\, \sin\theta) = (4\pi, 1)$ (very roughly the maximum coupling values consistent with unitarity). GECCO can probe this cross section for most masses and targets we consider.

Due to the annihilation cross section's scaling as the product of the couplings, each point in the $(m_\chi, \sv_{\bar{\chi}\chi,0})$ plane corresponds to a range of possible $\sin\theta$ values. The lower end of this range is determined by setting $g_{S\chi} \sim 4\pi$ while the upper end is $\sin\theta = 1$. We can conservatively map constraints on the Higgs portal model at each point in this plane by checking whether \emph{any} of the $\sin\theta$ values in this range are permitted. Applying this procedure using the constraints from~\cite{Krnjaic:2015mbs} leads to the orange region in the right panel of \cref{fig:higgs_portal}. At all points, these constraints are a few orders of magnitude more stringent than GECCO's discovery reach. This conclusion holds for other mediator masses $m_S > m_\chi$ above and below the resonance region around $m_S = 2 m_\chi$. We also plot an estimate of the CMB constraint assuming a kinetic decoupling temperature of $10^{-6} m_\chi$. While a more detailed calculation is possible, we do not pursue it here since the possibility of GECCO observing gamma-ray signals in this scenario is already strongly excluded by other constraints.

To guide the eye, we also plot curves corresponding to values of the coupling that give the correct DM relic abundance. GECCO can discover this benchmark Higgs portal model when the mediator is lighter than the DM and decays into photons or electrons, depending on the observing region. For both DM-mediator mass ratios shown, the process relevant for the standard relic abundance calculation is $\bar{\chi}\chi \to S S$. While this is not kinematically permitted for $m_\chi < m_S$ when the DM is nonrelativistic, it contributes dominantly to the thermal average involved in the relic abundance calculation since annihilations into SM final states are Yukawa-suppressed, making this an example of forbidden DM~\cite{DAgnolo:2015ujb}. Translating the value of $g_{S\chi}$ that gives the correct relic abundance for this scenario into $\sv_{\bar{\chi}\chi,0}$ additionally requires fixing $\sin\theta$, which we set to $1$ in the right panel of \cref{fig:higgs_portal}.\footnote{
    Note that there is a weak lower bound on $\sin\theta$ coming from requiring that the DM and mediator thermalize with the SM bath at early times.
} If the DM freezes out purely through annihilations into SM particles (as is the case for $m_S \gg m_\chi$), nonperturbatively large values of the DM-mediator coupling are required to give the correct relic abundance ($g_{S\chi} \gtrsim 100$), even for $\sin\theta = 1$. 

Given that we do not know the thermal history of the universe before big bang nucleosynthesis (BBN), the thermal relic cross sections we show can be evaded. For example, if the DM freezes out over-abundantly before BBN ($m_\chi / 20 \gtrsim T_\mathrm{BBN} \sim 1\, \mathrm{MeV}$), its density can be diluted through mechanisms like entropy injection into the SM bath via the decay of another heavy particle~\cite{Fornengo:2002db,Gelmini:2006pq,Hardy:2018bph} or late-time inflation~\cite{Davoudiasl:2015vba}, which have been explored carefully in the context of weakly interacting massive particle DM. For DM whose thermal relic density is lower than the observed cosmological dark matter density, the dark matter density can be increased through e.g. introducing a field that redshifts faster than radiation and dominates the universe's energy density at early times~\cite{Profumo:2003hq,relentlessdm}, or via non-thermal production. Detailed study of various ways of sidestepping the standard relic abundance constraints as well as a full relic abundance calculation that tracks the population of mediators falls outside the scope of this work.

\subsection{Model Example: Dark Photon}
\label{sec:vector}

Our vector-portal model is the well-known ``dark photon'' model in which we add a new \(\mathrm{U}(1)_{\mathrm{D}}\) gauge group and charge the DM under this group. We connect the dark sector and SM sector by letting the \(\mathrm{U}(1)_{\mathrm{D}}\) gauge boson mix with the Standard Model photon through \(\frac{\epsilon}{2}V_{\mu\nu}F^{\mu\nu}\) where \(\epsilon\) is a small mixing parameter and \(V^{\mu\nu}\) and \(F^{\mu\nu}\) are the dark photon and SM photon field strength tensors. The Lagrangian density is:
\begin{align}
    \mathcal{L} &= \mathcal{L}_{\mathrm{SM}} - \frac{1}{4}V_{\mu\nu}V^{\mu\nu}
    +\frac{\epsilon}{2}V_{\mu\nu}F^{\mu\nu} + 
    \bar{\chi}\qty(i\cancel{\partial}-m_{\chi})\chi + 
    g_{\chi V}V_{\mu}\bar{\chi}\gamma^{\mu}\chi
\end{align}
where \(V_{\mu}\) is the dark-photon. The kinetic terms for the 
\(\mathrm{U}(1)\) fields are diagonalized by shifting the SM-photon field by \(A_{\mu} \to A_{\mu} + \epsilon V_{\mu}\) and ignoring terms \(\order{\epsilon^2}\). The result is that all electrically-charged SM fields receive a small dark charge and the DM receives a small electric charge. After integrating out the heavy SM field and matching onto the chiral Lagrangian, we end up with the following interaction Lagrangian between the dark photon
and the light SM fields and meson:
\begin{widetext}
\begin{align}
    \mathcal{L}_{V-\mathrm{SM}} &= 
    - e V_{\mu}\sum_{\ell}\bar{\ell}\gamma^{\mu}\ell
    + i\epsilon e V_{\mu}\qty[\pi^{-}\partial_{\mu}\pi^{+} -\pi^{+}\partial_{\mu}\pi^{-}]
    -\frac{e^2}{32\pi^2}\epsilon^{\mu\nu\alpha\beta}F_{\mu\nu}V_{\alpha\beta}\qty(\frac{\pi^{0}}{f_{\pi}})
\end{align}
\end{widetext}
where \(\ell\) is either the electron or muon. The first two terms come from the covariant derivatives of the leptons and charge pion. The last term is a shift in the neutral pion decay, stemming from the Wess-Zumino-Witten Lagrangian~\cite{wess1971consequences, witten1983global}.

In our analysis, we focus on the regime where the mediator is heavier than the dark matter mass, taking \(3 m_{\chi} = m_{V}\). With this choice, we are able to recycle previously studied constraints produced by non-astrophysical experiments. The strongest constraints on dark photon models for the masses we are interested in come from the \(B\)-factory BaBar~\cite{Lees_2017} and beam-dump experiments such as LSND~\cite{Aguilar_2001}. Studies using the datasets of these experiments were able to constraint the dark photon model by looking for the production of dark photons which then decay into dark matter (see, for example Ref.~\cite{batell2009exploring, lees2017search, 10.1093/ptep/ptz106}); in the case of BaBar, the relevant process is \(\Upsilon(2S),\Upsilon(3S)\to \gamma + V \to \gamma + \mathrm{invisible}\), while the relevant process for LSND is \(\pi^{0}\to\gamma + V \to \gamma + \mathrm{invisible}\). We adapt the constraints computed in Ref.~\cite{10.1093/ptep/ptz106} (see Fig.(201) for the constraints and the text and references therein for details).

\begin{figure}[t]
    \centering
    \includegraphics[width=\textwidth]{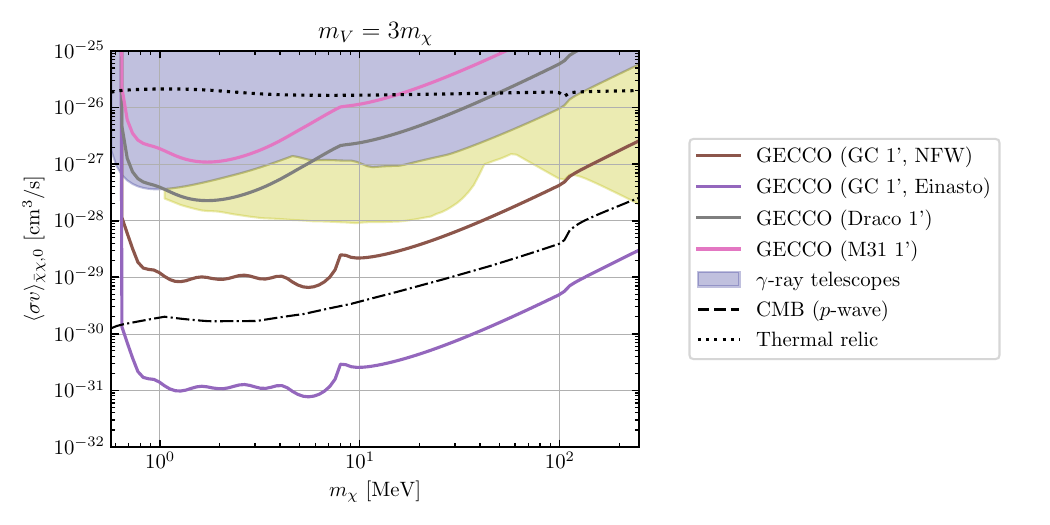}
    \caption{Projected constraints on the dark matter annihilation cross section for the dark photon model from GECCO (solid lines). 
    The blue shaded region shows the combined constraints from COMPTEL, EGRET, FERMI and INTEGRAL. The orange region shows the region excluded by BaBar and LSND. We show the contour yielding the correct dark matter relic density with the dotted black line.}
    \label{fig:km_constraint}
\end{figure}

In \cref{fig:km_constraint}, we show the combined constraints from BaBar and LSND in orange. As in \cref{sec:higgs_portal}, we show the constraints from existing gamma-ray telescope constraints (in blue), constraints from CMB (dashed black) and a contour where we find the correct relic density for the dark matter through standard thermal freeze-out through annihilation into Standard Model particles (dotted black). While our results show that the dark photon model is which dark matter is produced via standard thermal freeze out is already well excluded, we again point out that there are mechanisms for producing DM through nonthermal processes; see the end of the previous section for further discussion and Ref.~\cite{Evans:2019jcs} for a specific example using entropy dilution for a dark photon-mediated DM model. The projected constraints for GECCO for various targets and DM profiles are shown with solid lines. Our results demonstrate that GECCO's potential to significantly extend current constraints, and, more importantly, to offer opportunities for discovery of this class of well-motivated dark matter candidates.

\subsection{Model Example: Right-Handed Neutrino}\label{sec:rhn}

The decaying DM model we investigate is one in which the DM is given by a right-handed (RH) neutrino (i.e. a Weyl spinor transforming as a singlet under all Standard Model gauge groups) featuring a non-zero mixing with left-handed ``active'' neutrinos. We present the details of our RH neutrino model in App.~(\ref{app:rhn}). RH neutrinos are well-known and well-motivated DM candidates (for a recent review see e.g. Ref.~\cite{Boyarsky:2018tvu}). For the range of masses and lifetimes of interest here, the mixing angle must be extremely small: the two-body decay widths corresponding to a RH neutrino of mass $m_N$ with mixing angle with active neutrinos $\theta$ read
\begin{align}
    \Gamma(N_{a}\to\pi^{0} \nu_{b}) &= \delta_{ab}\frac{f_{\pi}^2G_F^2m_{N}^3 \theta^2}{8\pi}\qty( 1 - x_{\pi^{0}}^2)^2,\\
    %%%%%%%%%%%%%%
    \Gamma(N_{a}\to\pi^{\pm} \ell_{b}^{\mp}) 
    &= 
    \delta_{ab}\frac{f_{\pi}^2G_{F}^2|V_{ud}|^2\theta^2 m_{N}^3}{2\pi}
    \lambda^{1/2}(1,x_{\ell}^2,x_{\pi^{\pm}}^2)
    \qty[ \qty(1-x_{\ell}^2)^2-x_{\pi^{\pm}}^2\qty(1+x_{\ell}^2)],\\
    %%%%%%%%%%%%%%
    \Gamma(N_{a}\to\nu_{b} \gamma) 
    &\sim 
    \delta_{ab}\frac{9\alpha_{\mathrm{EM}}G_{F}^2m_{N}^5\theta^2}{64\pi^4},
\end{align}
while the three-body decay widths are
\begin{align}
    \Gamma(N_{a}\to\nu_{a}\nu_{b}\nu_{b}) 
    &\sim
    \qty(1 + \delta_{ab})
    \frac{K}{48},\\
    %%%%%%%%%%%%%%
    \Gamma(N_{a}\to\nu_{a} \ell_{b}^{+} \ell_{b}^{-}) &=
    \frac{K}{24}\bigg{[}
        c_{1,ab}((1 - 14 x^2 - 2 x^4 - 12 x^6) s(x) + 12 x^4 (x^4 - 1) \ell(x))\\
   &\hspace{0.5cm} 
   + c_{2,ab} (x^2 (2 + 10 x^2 - 12 x^4) s(x)
   + 6 x^4 (1 - 2 x^2 + 2 x^2) \ell(x))
   \bigg{]}\notag\\
    %%%%%%%%%%%%%%
    \Gamma(N_{a}\to\nu_{b} \ell_{a}^{+} \ell_{b}^{-}) &=
    \frac{K}{2}
    \int_{(x_{a}+x_{b})^2}^{1}\frac{\dd{x}}{x}
    \qty(x-x_{a}^2 -x_{b}^2)\qty(1-x)^2\lambda^{1/2}(x,x_a^2,x_b^2), 
    \quad\qty(a\neq b)
    \\
    %%%%%%%%%%%%%%
    \Gamma(N_{a}\to\nu_{a} \pi^{+} \pi^{-}) &=
    \frac{K}{96}
    \qty(1-2s_{W}^2)^2
    \int_{4x_{\pi}^2}^{1}\dd{z}
    \qty(1-z)^2\qty(1+2z)\beta^{3}(m_{N}^2z)\\
    %%%%%%%%%%%%%%
    \Gamma(N_{a}\to\ell^{\mp}_{a} \pi^0 \pi^{\pm}) &=
    \frac{K}{48}
    \abs{V_{ud}}^2
    \int_{4x_{\pi}^2}^{(1-x_{a})^{2}}\dd{z}
    \qty(\qty(1-x_a^2)^2 + z\qty(1+x_a^2) - 2 z^2)\\
    &\hspace{4cm}\times
    \lambda^{1/2}(1,z,x_a^2)\beta^{3}_{\pi}(m_{N}^2z)\notag
    %%%%%%%%%%%%%%
\end{align}
where \(K = G_{F}^{2}m_{N}^{5}/\pi^{3}\), \(s_{W}\) is the sine of the weak mixing angle, \(\lambda(a,b,c)=a^2+b^2+c^2-2ab-2ac-2bc\), \(x_{X} = m_{X}/m_{N}\), \(x_{a,b} = m_{\ell_{a,b}}/m_{N}\), \(\beta_{\pi}(s)=\sqrt{1-4m_{\pi}^2/s}\), \(s(x) = \sqrt{1-4x^2}\) and 
\(\ell(x) = \log(1/(x^2(1+s(x))))\). 
The constants \(c_{1,ab}\) and \(c_{2,ab}\) are
\begin{align}
    c_{1,ab} &= \frac{1}{4}\qty(1+4s_{W}^2+8s_{W}^4), &
    c_{2,ab} &= 2\qty(2s_{W}^2 + 1), & &\qty(a= b)\\
    c_{1,ab} &= \frac{1}{4}\qty(1-4s_{W}^2+8s_{W}^4), &
    c_{2,ab} &= 2\qty(2s_{W}^2 - 1), & &\qty(a\neq b)
\end{align}

For RH neutrino masses below the pion threshold, \(N\to\nu\ell\ell\) and \(N\to\nu\nu\nu\) decay modes are dominant. In this regime, photons are produced via the one-loop decay of the RH-neutrino into \(\nu \gamma\) and through radiation off a charged lepton, if \(N\to\nu \ell^{+} \ell^{-}\) is kinematically accessible. Once the pion threshold is crossed, the two body finals states \(N\to\pi^0 \nu_{\ell}\) and \(N\to\pi^{\pm} \nu_{\ell}^{\mp}\) dominate and photons are produced via the decay of pions and radiation off charged states.

\begin{figure*}[ht!]
    \centering
    \includegraphics[width=\textwidth]{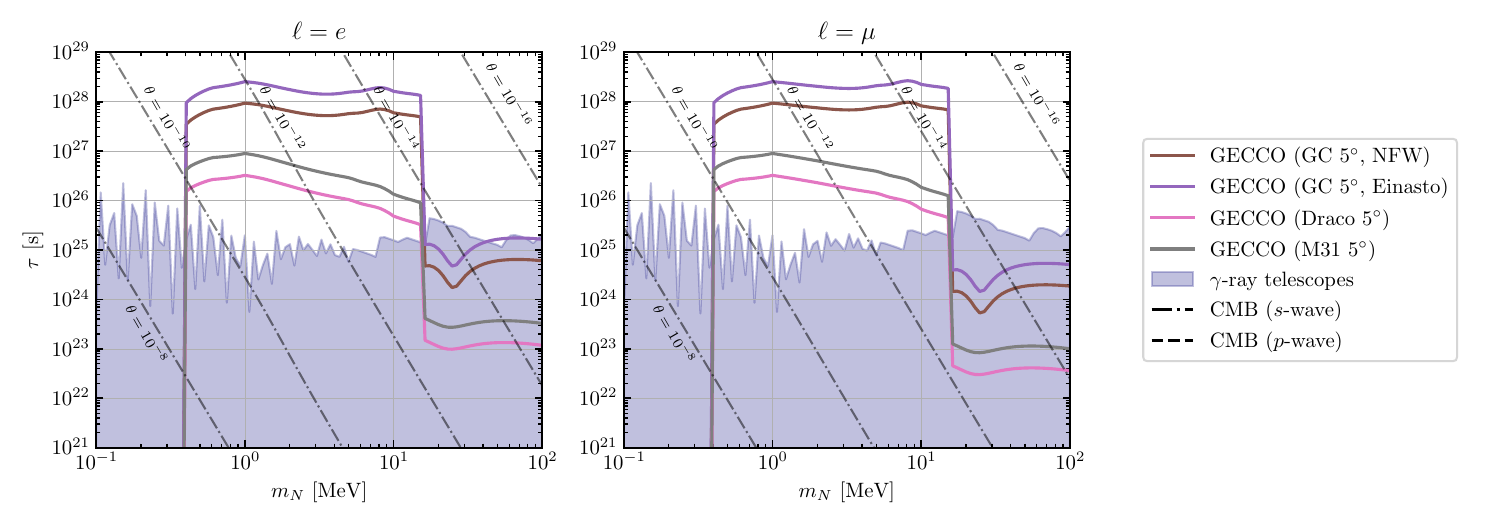}
    \caption{Projected constraints on the RH-neutrino lifetime (solid lines). 
    The area shaded in light blue is excluded by current observations, as in the previous plots. We also show, with dot-dashed contours, the mixing angle corresponding to parameter space shown in the figure.}\label{fig:rhn}
\end{figure*}

We show contours of constant $\theta$ on the lifetime versus mass plot in \cref{fig:rhn}. We do not assume here any specific RH neutrino production mechanism in the early universe. In the mass range of interest, the most natural, although by all means not the only, scenario is non-thermal production from the decay of a heavy species $\phi$ coupled to the RH neutrino via a Yukawa term of the form $y\phi\bar N N$ (see e.g. Ref.~\cite{Shaposhnikov_2006}). The yield depends on a variety of assumptions, including whether the $\phi$ is in thermal equilibrium or not,  which other decay channels it possesses, and the number of degrees of freedom that populate the universe as a function of time/temperature. However, production of RH neutrinos with the right abundance is generically possible across the parameter space we show in \cref{fig:rhn}.

The phenomenological constraints for RH neutrinos are weak for the masses and mixing angles of interest here. We refer the Reader to fig.~4 of Ref.~\cite{deGouvea:2015euy} for an extensive review. In short, the most stringent constraints occur for mixing with the electron-type active neutrino,  for a non-trivial CP phase and lepton-flavor violation structure. The strongest constraints, from neutrino-less double-beta decay, do not constrain values of the mixing angle to be smaller than $\theta\sim 10^{-8}$, even in the most favorable case. In the case of muon mixing, at or below 100 MeV the constraints are never stronger than $\theta\sim 10^{-4}$. Finally, in the weakest constraints case, that with tau neutrino mixing, the constraints on the mixing angle occur only for $\theta\gtrsim 10^{-2}$. We conclude that there are essentially no meaningful phenomenological constraints on the parameter space shown in \cref{fig:rhn}, in contrast to the situation for $\order{\mathrm{keV}}$-scale sterile neutrinos (see e.g. Ref.~\cite{Boyarsky:2018tvu}).

Our results in \cref{fig:rhn} indicate that a signal from sterile neutrino dark matter decay will be detectable from the Galactic Center over a wide range of masses and lifetimes. Limits will improve, for RH neutrinos in the few hundreds of keV range, by up to three orders of magnitude. A signal will also possibly be detectable for masses up to 100 MeV, and from targets different from the Galactic Center, such as M31 and Draco, for short enough lifetimes. Constraints from CMB observations are negligible~\cite{Essig:2013goa}.

We would like to point out a couple of features present in \cref{fig:rhn}). First, around $m_{N}\sim$ 10 MeV the GECCO constraint drops due to the fact that center of the gamma-ray line from $N\to\nu\gamma$ moves outside GECCO's sensitivity. The drop is more substantial in the case of muon mixing since there is a larger branching fraction into $\nu\gamma$ (in the electron mixing case, the branching fraction to $\nu\gamma$ is suppressed due to an enhanced branching fraction to $\nu_{e} e^{\mp}e^{\pm}$). Additionally, in the case where the RH neutrino mixes with the muon neutrino, we find a dip around $m_{N}\sim m_{\pi^{0}}$ because of the opening of the $N\to\nu_{\mu}\pi^{0}$ channel, which dominates the decay width of the RH neutrino. The photon spectrum from $N\to\nu_{\mu}\pi^{0}$ produces a box from $\pi^{0}\to\gamma\gamma$ with a width equal to the pion momentum. Thus, near $m_{N}\gtrsim m_{\pi^{0}}$, the box is narrow and outside GECCO's sensitivity. Once the $N\to\mu^{\mp}\pi^{\pm}$ channel opens up ($m_{N}>m_{\mu}+m_{\pi^{\pm}}$), a continuum spectrum is produced and the constraints increase. Note that this dip is not visible in the case where the RH neutrino mixes with the electron neutrino since $N\to e^{\mp}\pi^{\pm}$ opens up closely after $N\to \nu_{e}\pi^{0}$.

\section{Searches for Light Primordial Black Hole Evaporation}
\label{sec:pbh}

The discovery of gravitational radiation from binary black hole mergers ushered a renewed interest in black holes of primordial rather than stellar origin as dark matter candidates (for recent reviews, see e.g. Refs.~\cite{Green:2020jor,Carr:2020gox}). In a recent study, we considered Hawking evaporation from primordial black holes with lifetimes on the order of the age of the universe to $10^6$ times the age of the universe~\cite{Coogan:2020tuf}. There we corrected shortcomings of similar past analysis pertaining to the treatment of final state radiation and to the extrapolation of hadronization results outside proper energy ranges. We carried out a complete calculation of particle emission for Hawking temperatures in the MeV, and of the resulting gamma-ray and electron-positron spectrum.

Our key finding is that MeV gamma-ray telescopes are ideally poised to potentially discover Hawking radiation from light but sufficiently long-lived primordial black holes, specifically in the mass range between $10^{16}$ and $5 \times 10^{17}$ grams. The Hawking temperature scales with the holes' mass as $T_H\approx (10^{16} \ {\rm g}/M)\ {\rm MeV}$. As a result, especially towards the more massive end of that mass range, the bulk of the emission stems from prompt primary photon emission at higher energy, and from secondary emission from electrons at lower energy. 

\begin{figure}
    \centering
    \includegraphics[width=0.5\textwidth]{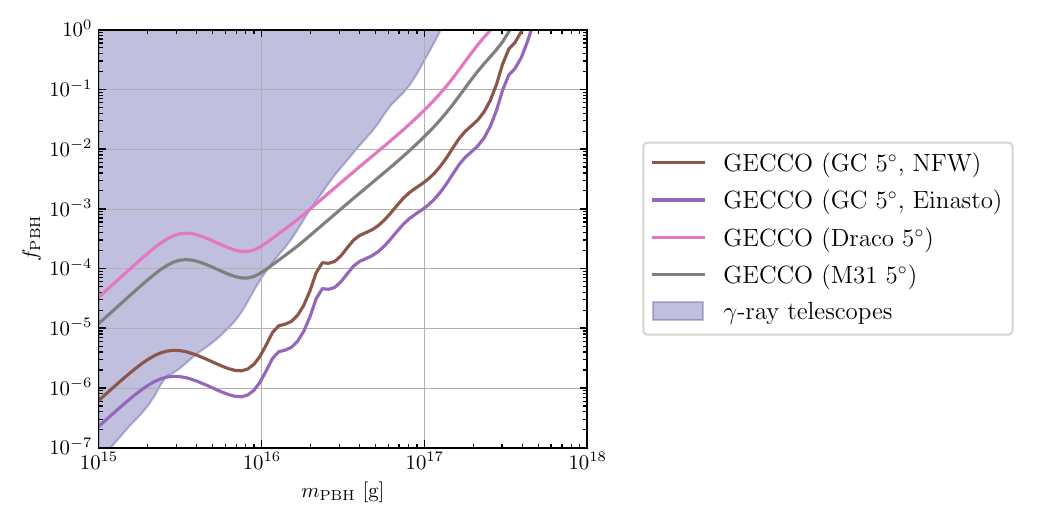}
    \caption{GECCO's $5\sigma$ discovery reach for detecting Hawking radiation from evaporating primordial black hole dark matter (solid lines). 
    The blue region shows existing constraints, the strongest of which comes from COMPTEL data~\cite{Coogan:2020tuf}. We assume a monochromatic mass function.}
    \label{fig:pbh_limits}
\end{figure}

Emission from the central region of the Galaxy and from nearby astrophysical systems with significant amounts of dark matter can be detectable with GECCO, as we show here. The calculation of the flux from black hole evaporation is as follows: a non-rotating black hole with mass $M$ and corresponding Hawking temperature $T_H=1/(4\pi G_N M)\simeq 1.06 (10^{16}\ {\rm g}/M)\ {\rm MeV}$, with $G_N$ Newton's gravitational constant, emits a differential flux of particles per unit time and energy given by 
\begin{equation}\label{eqn:hawkrad}
    \pdv{N_i}{E_i}{t} = \frac{1}{2\pi} \frac{\Gamma_i(E_i, M)}{e^{E_i/T_H}-(-1)^{2s}},
\end{equation}
where $\Gamma_i$ is the species-dependent grey-body factor, and $E_i$ indicates the energy of the emitted particle of species $i$. Unstable particles decay and produce stable {\em secondary} particles, including photons. The resulting differential photon flux per solid angle from a region parameterized by an angular direction $\psi$ is obtained by summing the photon yield $N_\gamma$ from all particle species the hole evaporates to:
\begin{equation}
    \dv{\phi_\gamma}{E_\gamma} = \frac{1}{4\pi \, M} \int_{\mathrm{LOS}} \dd{l} \rho_{\rm DM}(l, \psi) \, f_{\rm PBH} \, \pdv{N_\gamma}{E}{t}.
\end{equation}
Notice that upon integrating over the appropriate solid angle this expression is analogous to the one for the gamma-ray flux from decaying DM, containing the same $D$ factor (c.f. \cref{eq:flux}).

As for the calculation of the grey-body factors, we employ the publicly available code {\tt BlackHawk}~\cite{arbey2019BlackHawk}. {\tt BlackHawk} provides primary spectra of photons, electrons and muons. We then model the final-state radiation off the charged final state particles by convolving the primary particle spectrum with the Altarelli-Parisi splitting functions at leading order in the electromagnetic fine-structure constant $\alpha_{\mathrm{EM}}$~\cite{Chen:2016wkt,Altarelli:1977zs}. For the unstable particles, such as pions, we use {\tt hazma} to compute the photon spectrum from decays. The total resulting photon spectrum is then given by:
\begin{align}
    &\pdv{N_{\gamma}}{E_{\gamma}}{t} = \pdv{N_{\gamma,\mathrm{primary}}}{E_{\gamma}}{t}\\ 
    &\quad + 
    \sum_{i=e^{\pm},\mu^{\pm},\pi^{\pm}}\int\dd{E_{i}}
    \pdv{N_{i,\mathrm{primary}}}{E_{i}}{t}
    \dv{N^{\mathrm{FSR}}_{i}}{E_{\gamma}}\notag\\
    &\quad + \sum_{i=\mu^{\pm},\pi^{0},\pi^{\pm}}\int\dd{E_{i}}
    \pdv{N_{i,\mathrm{primary}}}{E_{i}}{t}
    \dv{N^{\mathrm{decay}}_{i}}{E_{\gamma}},\notag
\end{align}
where the FSR spectra are given by:
\begin{align}
    \dv{N^{\mathrm{FSR}}_{i}}{E_{\gamma}}&= \dfrac{\alpha_{\mathrm{EM}}}{\pi Q_f}P_{i\to i\gamma}(x)\qty[\log(\dfrac{(1-x)}{\mu_{i}^2})-1]\notag,\\
    P_{i\to\gamma i}(x)& = \begin{cases}
        \frac{2(1-x)}{x}, & i=\pi^{\pm}\\
        \frac{1+(1-x)^2}{x}, & i=\mu^{\pm},e^{\pm}\\
    \end{cases},
\end{align}
with $x=2E_{\gamma}/Q_f$, $\mu_{i}=m_{i}/Q_f$ and $Q_f=2E_f$. We give for explicit expressions of $dN^{\mathrm{decay}}/dE_{\gamma}$ for the muon, neutral and charged pions in Ref.~\cite{hazma}.

In evaluating GECCO's discovery reach we consider the same targets as in the preceding section: the Galactic Center with an NFW and an Einasto dark matter density profile, M31, and Draco. Assumptions on observing time are identical as before, and we use the same procedures to set limits and make projections as described in \cref{sec:id-procedure}. We additionally refer to our study of the discovery prospects of several proposed MeV gamma-ray telescopes for further details~\cite{Coogan:2020tuf}. The strongest existing bounds on evaporating PBHs were derived in that work using COMPTEL data \cite{Kappadath:1993}. Other competitive constraints come from INTEGRAL~\cite{Laha:2020ivk}, CMB data~\cite{Clark:2016nst,Poulin:2016anj}, EDGES 21 cm observations~\cite{Clark:2018ghm}, Voyager 1 $e^\pm$ measurements~\cite{Boudaud:2018hqb}, the 511 keV line~\cite{DeRocco:2019fjq,Laha:2019ssq}, dwarf galaxy heating~\cite{Kim:2020ngi} and the extragalactic gamma-ray background measurements~\cite{Carr:2009jm}. We note that for large PBH masses the constraints from the extragalactic gamma-ray background measurements~\cite{Carr:2009jm} and INTEGRAL~\cite{Laha:2020ivk} outperform those from COMPTEL.

In summary, we show in \cref{fig:pbh_limits} that GECCO will offer the exciting possibility of directly detecting Hawking evaporation from primordial black holes, for instance if these objects constitute at least 0.001\% of the dark matter and have a mass of $10^{16}$ grams, or if they are a larger fraction of the dark matter and a mass up to $5\times 10^{17}$ grams. Under optimistic circumstances (e.g. the black holes weigh around $10^{17}$ grams and they are more than 10\% of the dark matter), GECCO will detect Hawking evaporation from multiple targets besides the Galactic Center, such as from nearby dSph (e.g. Draco) and galaxies (e.g. M31). This reach in PBH mass is an order-of-magnitude improvement over existing bounds.

%%%%%%%%%%%%%%%%%%%%%%%%%%%%%%%%%%%%%%%%%%%%%%%%%%%%%%%%%%%
\section{Exploring the Origin of the 511 keV Line}
\label{sec:511}

The discovery of 511 keV line emission from positron-electron pair annihilation in the central region of the Galaxy dates back to balloon-borne experiments since the 1970s (see e.g. Ref.~\cite{1975ApJ...201..593H}). Space telescopes, specifically OSSE on the Compton Gamma-Ray Observatory~\cite{Skibo:1997yk} and, more recently, the SPI spectrometer~\cite{Weidenspointner:2008zz, 2003A&A...407L..55J} and the IBIS imager on board INTEGRAL~\cite{DeCesare:2005du} have significantly increased the amount of information about the 511 keV emission. The overall intensity of the line is around $10^{-3}$ photons cm$^{-2}$ s$^{-1}$, and it originates from a region of approximately 10$^\circ$ radius around the Galactic Center. The emission does not appear to have any significant time variability, and its spatial smoothness, combined with the point-source sensitivity of the IBIS imager, places a lower limit of at least eight discrete sources contributing to the signal~\cite{2003A&A...407L..55J}.

Measurements of the diffuse emission at energies below and above 511 keV constrain the injection energy of the positrons and the properties of the medium where injection and annihilation occur. Most notably for constructing new physics interpretations of the signal, the absence of significant emission at energies higher than 511 keV indicates that the positron injection energy is bounded from above in the few MeV (at most $4 - 8.5\, \mathrm{MeV}$, allowing for a partially ionized medium~\cite{sizun2006, sizun2007}). In turn this implies an upper limit of around 3 MeV on the mass of putative dark matter particles annihilating to electrons and positrons in a neutral medium~\cite{1981SvAL....7..395A, beacom2006}. In absence of large-scale magnetic fields~\cite{prantzos2006}, the injection sources of positrons are constrained to lie within approximately 250 pc of the annihilation sites~\cite{jean2006}, thus indicating that the source distribution is quite close to the actual signal distribution in the sky~\cite{churazov2005, jean2006}.

The origin of the positrons in the Galactic Center is still actively debated. Morphological information, and the mentioned lower limit on the number of contributing sources, rules out as major contributors (although it does not rule out as co-contributors) single sources such as Sgr A*~\cite{1982AIPC...83..148L} or a single injection event such as a gamma-ray burst or a hypernova in the Galactic Center~\cite{1984AIPC..115..558L}. The bulk of the signal is however slated to originate from a distributed population of several sources that could not be resolved as individual point sources in prior observations~\cite{2005A&A...441..513K}.

Much enthusiasm surrounded the possibility that the 511 keV line originate from sources associated with new physics. Of these, the simplest possibility is perhaps the pair-annihilation of MeV-scale dark matter particles~\cite{2004PhRvL..92j1301B}. Other proposed scenarios include the decay of new particles such as sterile neutrinos~\cite{2005PhLB..605...15P}, axions~\cite{2004PhRvD..70f3506H}, neutralinos~\cite{2006MPLA...21..457B}, Q-balls~\cite{2005PhRvD..72h5015K}, mirror matter~\cite{2005IJMPD..14..143F}, moduli~\cite{2005PhLB..624..162K}, cosmic strings~\cite{2005PhRvL..95z1302F}, superconducting quark matter~\cite{2005PhRvL..94j1301O}, MeV-scale excitations of more massive particles~\cite{Finkbeiner:2007kk, 2007PhLB..651..208P}, or small accreting black holes~\cite{2006ApJ...641..293T}. The common denominator of all these ``exotic'' scenarios is a genuinely diffuse emission: the significant detection of point sources at 511 keV would robustly rule out a new physics origin for the signal. Here, we point out that GECCO's outstanding point source sensitivity would provide an exceptional probe to discriminate between an exotic and a conventional astrophysical origin for the signal.

A variety of conventional astrophysical sources have been considered for the production of positrons in the Galaxy contributing to the 511 keV signal. These include massive stars, pulsars as well as millisecond pulsars, core-collapse supernovae and SNe Ia, Wolf-Rayet stars, and low-mass X-ray binaries (LMXB), especially microquasars~\cite{Siegert:2015knp, Bandyopadhyay:2008ts}. In many instances, these astrophysical objects are also found much closer to the solar system than in the Galactic Center region. For instance, the closest Wolf-Rayet star, in the Gamma Velorum system, is around 350 pc away~\cite{Millour_2006}; the catalogue in Ref.~\cite{Liu:2007ts} includes an LMXB at a distance of 0.42 kpc (4U 1700+24) as well as at least four candidates closer than 2 kpc. The ATNF catalogue~\cite{Manchester:2004bp} contains several MSPs closer than 0.2 kpc, including J0437-4715 whose distance is 0.16 kpc (see also Ref.~\cite{Bogdanov:2012md}), J0605+3757 at 0.21 kpc, J0636+5129 and J1737-0811 also at 0.21 kpc, J2322-2650 at 0.23 kpc, J1017-7156 at 0.26 kpc, and J1400-1431 at 0.28 kpc. 

\begin{figure*}
    \centering
    \includegraphics[width=\textwidth]{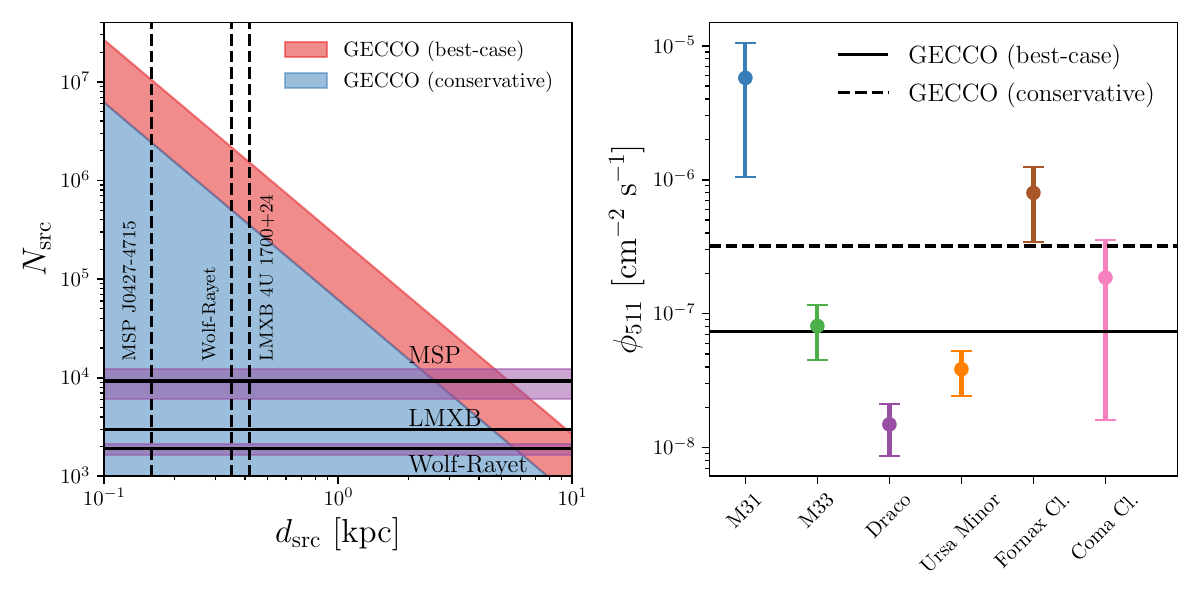}
    \caption{Left: the GECCO sensitivity to 511 keV individual point source on the plane defined by the number of sources contributing to the signal at the Galactic Center (assumed to all contribute the same 511 keV luminosity), versus the distance of the closest such source; we also indicate with vertical dashed lines the distance to the closest MSP, Wolf-Rayet star, and LMXB, and with horizontal dark green bands the estimates for the total number of MSP and Wolf-Rayet stars potentially contributing to the signal. Right: predictions for the 511 keV flux from a variety of nearby astrophysical objects, based on a signal scaling proportional to mass over distance squared. The horizontal dashed and solid lines correspond to GECCO's point source sensitivity best-case and conservative case.}
    \label{fig:511}
\end{figure*}

GECCO's angular resolution and point-source sensitivity make it ideally suited to enable to differentiate between a multiple discrete point sources versus a genuinely diffuse origin for the 511 keV emission. Specifically, if one source class dominated the positron emission, GECCO has a distinct chance to detect nearby members of that source class. To clarify and quantify this statement, we assume for simplicity that the 511 keV signal originates from $N_{\rm src}$ sources each with a luminosity $L_{\rm src}$ at an average distance of 8.12 kpc. Given that the 511 keV signal is approximately $\phi_{511}\simeq 3\times 10^{-3}\, \Delta\Omega\, {\rm cm}^2\, {\rm s}^{-1}\, {\rm sr}^{-1}$ over an angular region of 10 degrees, i.e. $\Delta\Omega\simeq 0.1$ sr, the flux expected from a single source at a distance $d_{\rm src}$ reads
\begin{equation}
    \phi_{\rm src}=\frac{L_{\rm src}}{4\pi d_{\rm src}^2}\simeq \frac{\phi_{511}}{N_{\rm src}}\left( \frac{8.12\ {\rm kpc}}{d_{\rm src}} \right)^2.
\end{equation}
We can thus compare the narrow line flux sensitivity of GECCO, which in the best-case scenario is $7.4\times 10^{-8}\ {\rm cm}^{-2}\, {\rm s}^{-1}$ and in the worse case scenario $3.2\times 10^{-7}\ {\rm cm}^{-2}\, {\rm s}^{-1}$ with the flux expected for a given putative source class point source. Specifically, we calculate the GECCO sensitivity on the plane of $N_{\rm src}$ vs $d_{\rm src}$. In the plot shown in fig.~\ref{fig:511} we indicate with vertical lines the closest known WR star, LMXB, and MSP, and with a horizontal line an estimate for the number of LMXB that could be responsible for the 511 keV line according to Ref.~\cite{Bandyopadhyay:2008ts} ($N_\mathrm{LXMB} \simeq 3000$), the estimate in Ref.~\cite{Calore:2015bsx} for the number of MSP in the Galactic Center region ($N_{\rm MSP}\simeq (9.2\pm3.1)\times 10^3$) and an estimate for the total number of Wolf-Rayet stars in the Milky Way from Ref.~\cite{10.1093/mnras/stu2525} ($N_\mathrm{WR} \simeq 1900 \pm 250$).

The plot shows that GECCO's sensitivity should enable the detection of any positron source responsible for a significant fraction of the 511 keV signal closer than 4 kpc.

Additional information on the nature of the origin of the 511 keV signal from the Galactic Center will be provided by observations of nearby systems such as the Andromeda galaxy (M31), the Triangulum galaxy (M33), nearby clusters such as Fornax and Coma, and nearby satellite dwarf galaxies such as Draco and Ursa Minor~\cite{Wolf:2009tu}. 

The crudest estimate of the predicted 511 keV signal is a simple mass to distance-squared ratio, which we report in \cref{tab:511keV-brightness}. According to our predictions, the 511 keV signal from M31 should be detectable by GECCO, as should the signal from the Fornax and (although marginally) the Coma cluster. We predict that instead M33, and local dSph should not be bright enough at 511 keV to be detectable by GECCO. Notice that Integral/SPI already searched for a 511 keV line from Andromeda (M31), reporting an upper limit to the flux of $1\times 10^{-4}\, {\rm cm}^{-2}\, {\rm s}^{-1}$~\cite{Bandyopadhyay:2008ts}.

Notice that certain types of new physics explanations such as dark matter decay would follow a similar scaling. Other new physics explanations such as e.g. eXcited dark matter~\cite{Finkbeiner:2007kk} would not, a critical factor being the typical velocity dispersion in a given system: no signal at all would be predicted from e.g. small galaxies such as Draco or Ursa Minor. The predictions for galaxies versus clusters of galaxies would depend upon the details of the model, but generally scale similarly to what reported in \cref{tab:511keV-brightness}. 

We use the estimate of Ref.~\cite{li2016modelling} for the Milky Way bulge total mass, and the flux quoted in Ref.~\cite{Siegert:2015knp} for the 511 keV flux from the bulge. We take the value for the total dynamical mass of M31 from Ref.~\cite{Evans_2000}, while the distance is from Ref.~\cite{2006Ap.....49....3K}; the total mass of M33 is from Ref.~\cite{1970A&A.....9..350B} and the distance from Ref.~\cite{2006Ap&SS.304..207B}. For the dSph we take data from Ref.~\cite{Wolf:2009tu}. Data for the Fornax cluster are from Ref.~\cite{Jordan_2007}, while for the Coma cluster from Ref.~\cite{1989ApJS...70....1A} and Ref.~\cite{Gavazzi_2009}. We propagate errors including those on masses, distances, and the observed 511 keV flux, and show our results in the right panel of \cref{fig:511}.

\begin{table}[]
    \centering
    \begin{tabular}{c c c c}
        \toprule
        Target & Mass [$M_\odot$] & Distance [kpc] & $\phi_{511}$ [cm$^{-2}$ s$^{-1}$]\\
        \midrule
        Milky Way &  $(1.69\pm0.12)\times 10^{10}$ & 8.5 & $(9.6\pm 0.7)\times 10^{-4}$\\
        \midrule
        M31 & $(8.5\pm 5)\times 10^{11}$  & $778\pm33$ & $(5.76 \pm 4.71)\times10^{-6}$ \\
        M33 & $(1.75\pm0.25)\times 10^{10}$ & $942\pm 73$ & $(8.09 \pm 3.58)\times 10^{-8}$ \\
        \midrule
        Draco & $(2.1\pm0.3)\times 10^7$ & $76\pm5$ & $(1.49 \pm 0.62)\times 10^{-8}$ \\
        Ursa Minor & $(5.6\pm0.7)\times 10^7$  & $77\pm4$ & $(3.85 \pm 1.44)\times 10^{-8}$ \\
        \midrule
        Fornax Cl. & $(7\pm2)\times 10^{13}$ & $(18.97\pm1.33)\times 10^3$ & $(7.98 \pm 4.55)\times 10^{-7}$\\
        Coma Cl. & $(5.1\pm3.2)\times 10^{14}$ & $(106.1\pm7.5)\times 10^3$ & $(1.86 \pm 1.70)\times 10^{-7}$\\
        \bottomrule
    \end{tabular}
    \caption{Predicted brightness of a 511 keV signal assuming a scaling proportional to mass over distance squared for a variety of astrophysical targets (see main text for references to the quoted masses, distances, and fluxes).}
    \label{tab:511keV-brightness}
\end{table}

%%%%%%%%%%%%%%%%%%%%%%%%%%%%%%%%%%%%%%%%%%%%%%%%%%%%%%%%%%%
\section{Discussion and Conclusions}
\label{sec:conclusions}

We explored and elucidated the scientific portfolio that would be enabled by the deployment of the proposed mid-scale Explorer class NASA mission GECCO as it pertains to dark matter and new physics. GECCO is ideally suited to explore MeV dark matter candidates as long as they decay and/or pair-annihilate. The new instrument would unveil dark matter signals up to four orders of magnitude fainter, for certain dark matter particle models, than the current observational sensitivity, and would make it possible to detect a dark matter signal from multiple astrophysical targets, reducing the intrinsic background and systematic effects that could otherwise obscure a conclusive discovery.

GECCO would enable the exciting possible direct detection of Hawking evaporation from primordial black holes with masses in the $10^{16}-5\times 10^{17}$ grams range, if they constitute a sizable fraction of the cosmological dark matter. Under favorable circumstances, GECCO might detect Hawking evaporation from more than one astrophysical target as well.

Finally, we showed the potential of GECCO to elucidate the nature of the 511 keV line, by virtue of its unprecedented line sensitivity and point-source angular resolution. We found that GECCO should be able to observe a 511 keV line from a variety of extra-Galactic targets, such as nearby clusters and massive galaxies and, potentially, even from nearby dwarf galaxies; in addition, GECCO should be able to detect single sources of the 511 keV emission, as long as they are reasonably close.

In summary, we have shown that GECCO would push the observational frontier of MeV gamma rays in ways that would enormously benefit the quest for fundamental questions in cosmology and particle physics, chiefly the nature and particle properties of the cosmological dark matter, and the origin of the mysterious 511 keV line emission from the center of the Galaxy.

\section*{Acknowledgements}
This work is partly supported by the U.S.\ Department of Energy grant number de-sc0010107. A.C. received funding from the Netherlands eScience Center (grant number ETEC.2019.018) and the Schmidt Futures Foundation.
A. Moiseev's effort is supported by NASA awards 80GSFC21M0002 and 80NSSC20K0573.

\appendix

\section{Right Handed Neutrino Dark Matter}\label{app:rhn}

In this appendix, we detail our model for right-handed neutrino (RHN) dark matter.
We consider the case with a single additional Majorana fermion that is neutral under all SM gauge groups. We take the Lagrangian density to be
\begin{align}\label{eqn:app_a_rhn}
    \mathcal{L} &= \mathcal{L}_{\mathrm{SM}}+ 
    i\hat{N}^{\dagger}\bar{\sigma}^{\mu}\partial_{\mu}N 
    -\dfrac{1}{2}\hat{m}_{N}\qty(\hat{N}\hat{N}+\hat{N}^{\dagger}\hat{N}^{\dagger})
    - y_{\ell}\qty(\hat{L}_{\ell}^{\dagger}\tilde{H}\hat{N}+\mathrm{h.c.}) 
\end{align}
where the RHN, \(N\), is the 2-component Weyl spinor and 
\(\hat{L}_{\ell} = \qty(\hat{\nu}_{\ell} \ \hat{e}_{\ell})^T\) with \(\ell\in\qty{e,\mu,\tau}\) represents the SM lepton doublets. For simplicity, we take the Yukawa coupling \(y_{\ell}\) to be non-zero for only a single generation, \(y_{k} = y\) and \(y_{\ell\neq k} = 0\).

For non-zero 
\(\hat{m}_N\), diagonalizing the neutrino mass matrix yields two majorana
spinors. The diagonalization can be performed by constructing a neutrino
mass matrix for the neutrino interactions states 
\(\vb*{\nu} = \mqty(\hat{\nu}_{\ell} \ \hat{N})\) and performing a Takagi 
diagonalization~\cite{dreiner2010two}. Explicitly, if the neutrino mass matrix is \(\bm{M}\), then it must be a complex symmetric matrix. A complex symmetric matrix is diagonalized through a unitary Takagi matrix \(\bm{\Omega}\), with \(\bm{\Omega}^{T}\bm{M}\bm{\Omega}\) resulting in the diagonal mass matrix.

For the Lagrangian density in Eq.~(\ref{eqn:app_a_rhn}), the unitary Takagi transformation matrix is:
\begin{align}
    \Omega &= \mqty(-i\cos\theta & \sin\theta\\ i\sin\theta &\cos\theta)
\end{align}
where \(\sin\theta\) is
\begin{align}
    \sin\theta &= \frac{vy}{\sqrt{2}\hat{m}_{N}} - \order{\frac{vy}{\hat{m}_{N}}}^3
\end{align}
and \(\cos\theta\sim 1\) at leading order.

To determine the interactions between the RHN and mesons, we begin by integrating out the electroweak states. The resulting effective interaction Lagrangian is described by the well-known 4-fermion Lagrangian:
\begin{align}
    \mathcal{L}_{\mathrm{N(int)}} = -\dfrac{4G_{F}}{\sqrt{2}}
    \eval{\qty[J_{\mu}^{+}J_{\mu}^{-} + \qty(J^{Z}_{\mu})^2]}_{\nu^{k}_{L}\to \sin\theta N -i\cos\theta\nu^{k}_{L}}
\end{align}
Where \(G_{F}\) is Fermi's constant and \(J^{\pm}_{\mu}\) and \(J^{Z}_{\mu}\) are the charged and neutral weak fermion currents, given by:
\begin{align}
    J^{+}_{\mu} &= \sum_{i}\bar{\nu}^{i}_{L}\gamma^{\mu}\ell^{i}_{L} + 
    \sum_{i,j}V^{\mathrm{CKM}}_{ij}\bar{u}^{i}_{L}\gamma^{\mu}d^{j}_{L}\\
    J^{Z}_{\mu} &= \dfrac{1}{2c_{W}}\sum_{i=1}^{3}
    \bigg{[}\qty(1-\frac{4}{3}s^2_{W})\bar{u}_i\gamma^{\mu}u_{i}
    +\qty(-1+\frac{2}{3}s^2_{W})\bar{d}_i\gamma^{\mu}d_{i}
                +\bar{\nu}^{i}_{L}\gamma^{\mu}\nu^{i}_{L}
                -\qty(1+2s^2_{W})\bar{\ell}^{i}\gamma^{\mu}\ell^{i}\notag
    \bigg{]}
\end{align}
with \(s_{W}\) and \(c_{W}\) being the sine and cosine of the weak mixing angle.

In order to calculate the interactions between the RH neutrino and mesons, we first determine the interaction Lagrangian written in terms of light quarks. Grouping the up, down and strange into a light-quark triplet \(\vb{q} = \qty(u \ d \ s)^T\), we can write the relevant interactions terms of the expanded 4-Fermi Lagrangian as:
\begin{align}\label{eqn:app_a_fourfermi}
    -\frac{\sqrt{2}}{4G_{f}}\mathcal{L}_{\mathrm{N(int)}} 
    &= 
    L^{+}_{\mu}L^{-}_{\mu} + (L^{0}_{\mu}+R^0_{\mu})^2 + 
    \bar{\vb{q}}\gamma^{\mu}\qty[2\vb{G}_{R}(L^{0}_{\mu}+R^{0}_{\mu})]P_{R}\vb{q} \\
    &\hspace{1cm}
    +\bar{\vb{q}}\gamma^{\mu}\qty[\qty(\vb{V}^{\dagger}L^{-}_{\mu} + \mathrm{h.c.}
    )+2\vb{G}_{L}\qty(L^{0}_{\mu}+R^{0}_{\mu})]P_{L}\vb{q} + 
    \cdots \notag
\end{align}
with the \(\cdots\) containing all terms without the RHN. The charged and neutral left and right handed currents which the light quarks interact with are given by:
\begin{align}
    R^{0}_{\mu} &= \frac{1}{2c_{W}}
    (i\cos\theta\bar{\nu}^{k}_{L}+\sin\theta \bar{N})\gamma^{\mu}
    (-i\cos\theta\nu^{k}_{L}+\sin\theta N) + \cdots\\
    L^{0}_{\mu} &= \frac{1}{2c_{W}}
    (i\cos\theta\bar{\nu}^{k}_{L}+\sin\theta \bar{N})\gamma^{\mu}
    (-i\cos\theta\nu^{k}_{L}+\sin\theta N) + \cdots\\
    L^{-}_{\mu} &= -i\cos\theta\nu^{k}_{L}\gamma^{\mu}\ell^{k}_{L} + 
    \sin\theta\bar{N}\gamma^{\mu}\ell^{k}_{L}+\cdots
\end{align}
\(\vb{G}_R\) and \(\vb{G}_L\) are the right and left light-quark coupling matrices to the \(Z\) boson, given by:
\begin{align}
    \vb{G}_{R} &= \frac{1}{2c_{W}}\mathrm{diag}\qty(1,-1,-1) + \vb{G}_{L}, &
    \vb{G}_{L} &= -\frac{s_{W}^2}{3c_{W}}\mathrm{diag}\qty(-2,1,1)
\end{align}
and \(\vb{V}\) is CKM coupling matrix for the light quarks:
\begin{align}
    \vb{V} &= \mqty(0 & V_{ud} & V_{us} \\ 0 & 0 & 0 \\  0 & 0 & 0)
\end{align}
With the interactive Lagrangian written in the form of Eq.~(\ref{eqn:app_a_fourfermi}), matching onto the chiral Lagrangian is straightforward. The terms of the from 
\(\bar{\vb{q}}\gamma_{\mu}J^{\mu}_{L,R}P_{L,R}\vb{q}\) are matched onto the ``covariant'' derivative of the meson matrix of the chiral Lagrangian while the terms without quarks are unaffected. The result is:
\begin{align}
    \mathcal{L} = \frac{f_{\pi}^2}{4}\Tr[\qty(D_{\mu}\vb{\Sigma})^{\dagger}\qty(D_{\mu}\vb{\Sigma})]
     + L^{+}_{\mu}L^{-}_{\mu} + \qty(L^{0}_{\mu} + R^0_{\mu})^2 + \cdots
\end{align}
where \(f_{\pi}\) is the pion decay constant \(f_{\pi}\sim 92\) MeV and the \(\vb{\Sigma}\) field is the pseudo-Goldstone matrix containing the meson made from \(u, d\) and \(s\) quarks:
\begin{align}
    \vb{\Sigma} = \mqty(\pi^{0} + \eta/\sqrt{3} & \sqrt{2}\pi^{+} & \sqrt{2}K^{+}\\
    \sqrt{2}\pi^{-} & -\pi^{0} + \eta/\sqrt{3} & \sqrt{2}K^{0}\\
    \sqrt{2}K^{-} & \sqrt{2}\bar{K}^{0} & -2\eta/\sqrt{3})
\end{align}
and the covariant derivative is:
\begin{align}
    D_{\mu}\vb{\Sigma} &= \partial_{\mu}\vb{\Sigma} -i r_{\mu}\vb{\Sigma} + i\vb{\Sigma}l_{\mu}\\
    r_{\mu} &= 2\vb{G}_{R}R^{0}_{\mu}\\
    l_{\mu} &= \qty(\vb{V}^{\dagger}L^{-}_{\mu} + \mathrm{h.c.})+2\vb{G}_{L}\qty(L^{0}_{\mu}+R^{0}_{\mu})
\end{align}

\bibliography{references}

\end{document}